# Upgraded Metallurgical Grade Silicon for solar electricity production: a comparative Life Cycle Assessment

_Laura Méndez[1]_, Eduardo Forniés[1], Daniel Garrain[2], Antonio Pérez Vázquez[3], Alejandro Souto[3], Timur Vlasenko[4]

[1]Aurinka PV Group, Marie Curie 19, 28521 Rivas-Vaciamadrid (Madrid), Spain
[2]CIEMAT, Av. Complutense 40, 28040 Madrid (Madrid), Spain
[3]Ferroglobe, Pol. Sabón, 15142 Arteixo (La Coruña), Spain
[4]FerroSolar, Ucrania 6, 13500 Puertollano (Ciudad Real), Spain

*Corresponding author: email: lmendez@aurinkapv.com*

**Abstract**

Solar grade silicon (SoG-Si) is a key material for the development of crystalline silicon photovoltaics (PV), which is expected to reach the tera-watt level in the next years and around 50TW in 2050. Upgraded metallurgical grade silicon (UMG-Si) has already demonstrated to be a viable alternative to standard polysilicon in terms of cost and quality. This study presents the life cycle assessment (LCA) of UMG obtained by the FerroSolar process. Moreover, it shows the environmental impacts of PV modules and electricity generation based on this material. For this, an exhaustive review of the life cycle inventory (LCI) of PV value chain, from metallurgical grade silicon (MG-Si) down to electricity generation, has been carried out updating inputs for all processes. The Balance of System (BoS) has also been updated with real state of the art data for a fixed open ground large PV site (100 MW). Two different electricity mixes, with low and high carbon intensities, have been considered. The results reveal that for PV electricity generation using UMG instead of polysilicon leads to an overall reduction of Climate change (CC) emissions of over 20%, along with an improvement of the Energy Payback Time (EPBT) of 25%, achieving significantly low values, 12 $gCO_{2eq}$ / $kWh_e$ and 0.52 years, respectively. Moreover, it is shown that UMG silicon feedstock is not the main contributor to the carbon and energy footprint of the produced electricity, leaving the first place to PV module manufacturing.

**Graphical Abstract**

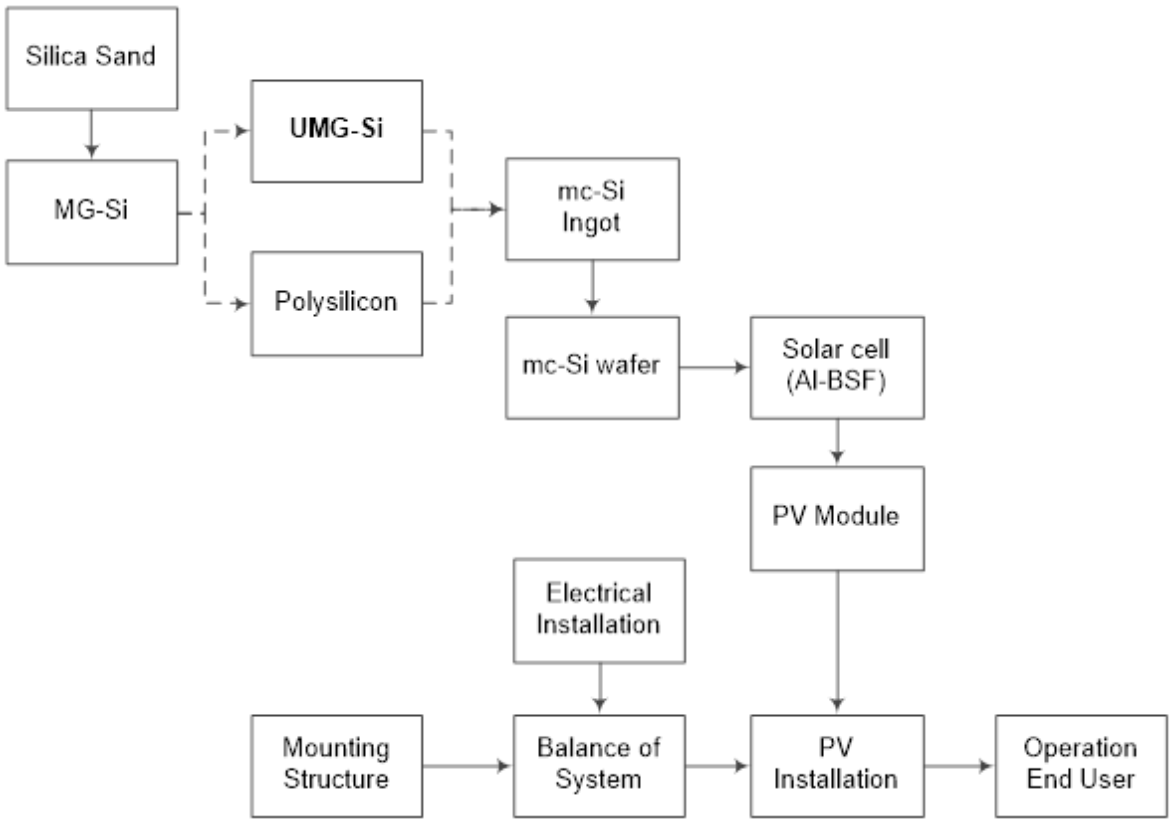

# 1 Introduction

Solar photovoltaics is a crucial technology for achieving a decarbonized electricity in the coming years (Breyer et al., 2018). The power sector is the main responsible of the world's greenhouse gases (GHG) emissions, with approximately 70%, due to the predominant share of fossil fuels. The decarbonization of the power sector is mandatory to achieve the objective of limiting the global average temperature rise to 1.5 ºC (Jäger-Waldau, 2019), as stablished in the Paris Agreement, back in 2016.

PV has already demonstrated to be an economically viable source of electricity. Its current and expected Levelized costs of Energy (LCOE) are below either fossil or other renewable energies (IEA, 2020; Vartiainen et al., 2020). To achieve the present status, many contributions have been made along the whole value chain in terms of cost reduction and increase of efficiency. Silicon material usage for crystalline cells has been reduced significantly during the last decade from around 16 to below 4 $g/W_{pk}$ due to increased efficiencies, thinner wafers and wires as well as larger ingots and cells (VDMA, 2019).

As PV worldwide installed capacity points to the terawatt-level in the next five years (Haegel et al., 2019), the production of silicon for PV applications will have to grow accordingly to cover the demand. Nowadays, crystalline silicon technology accounts for over 95% of the worldwide market and it can be safely assumed that will remain the same for the following years (Philipps and Warmuth, 2020).

Under the denomination of "solar grade silicon" (SoG-Si), different grades are described, regarding to their concentration of impurities according to the "*Specification for Virgin Silicon Feedstock Materials for Photovoltaic Applications*" (SEMI PV17-1012) (Ceccaroli et al., 2016). Nowadays the market demand of solar grade silicon is almost completely covered by polysilicon, produced by different configurations of the Siemens process. Alternatives to Siemens polysilicon are Fluidized Bed Reactor (FBR) Solar Silicon and upgraded metallurgical grade silicon (UMG), and even direct carbothermic reduction of silica. All of the have in common their lower energy consumption (Forniés et al., 2016; Maldonado, 2020), and therefore low energy and carbon footprints. Even though the penetration in the market of alternative feedstocks of has not yet become significant, these still raise the attention of the research community (Chen et al., 2019; Cherif et al., 2019; Du and Liu, 2020; Ye et al., 2019) and the industry (Osborne, 2020; Verdu, 2020).

The UMG silicon assessed in this work has been manufactured through the metallurgical route by means of the process developed by Ferrosolar in Spain. In a previous mass production test, performed in commercial solar cells and modules production lines, this feedstock has proven to be appropriate for PV applications (Forniés et al., 2019), reaching, in a conventional production line, up to 20.76% of solar cell efficiency with multicrystalline cells made of 100% UMG silicon. Additional results have been recently presented (Fornies et al., 2021), on defect engineering and outdoor degradation of UMG-Si PV modules compared to polysilicon, including some partial results presented in the present work.

For the past decade, new PV manufacturing capacity has been mostly installed in China and other places in Asia, covering the whole value chain, from polysilicon to PV modules. It has been previously pointed out that the environmental impact of these production sites is higher than in other regions, in spite of the contribution of higher scale to the optimization of the technologies (Leccisi et al., 2016; Stamford and Azapagic, 2018; Yue et al., 2014). In the case of the European Union, the attention has been drawn to the need of a larger local installed capacity of PV manufacturing in order to ensure the supply of PV devices for the predicted new installations (Jäger-Waldau et al., 2020), that must be cost-effective (Rentsch et al., 2019) in order to be successful.

The potential of PV as leading decarbonization technology rests upon the zero-emissions production of electricity for its lifetime, but the environmental impact associated to its full life cycle needs to be addressed. Countries as France, and recently South Korea, are regulating the carbon footprint of PV modules eligible for large scale tenders and subsidies (Stoker, 2020), as the impact of PV electricity is concentrated in the production stage.

Life Cycle Assessment is a comprehensive, standardized and internationally recognized approach for quantifying all emissions, resource consumption and related environmental and health impacts linked to a product. A fair amount of publications have dealt with LCA for PV for the past three decades including various recent reviews (Ludin et al., 2018; Muteri et al., 2020). Other studies show the relationship between cost and environmental impact (Louwen et al., 2016; Mayer et al., 2021), which are deeply related to Si material and energy consumption, and consequently to environmental footprint.

The objective of this research work is to assess the environmental impacts of UMG silicon based solar PV systems in comparison with traditional state of the art polysilicon-based ones, by means of a comparative LCA. Climate change (CC) emissions and EPBT have been selected as main indicators and Cumulative Energy Demand (CED) and Acidification (terrestrial and freshwater) (ATF) are also reported. To achieve a fair comparison of UMG and standard polysilicon, a comprehensive full PV value chain analysis has been carried out. Moreover, the analysis has been made for two electricity mixes, with different carbon intensities, as electrical power input has a decisive role in the overall impact of PV value chain (Yue et al., 2014).

## 2 Methods

The environmental impacts have been estimated using process-based LCA, according to the Methodology Guidelines on Life Cycle Assessment of Photovoltaic Electricity published by the International Energy Agency (Frischknecht et al., 2016) and following the Environmental Footprint (EF) 3.0 impact assessment method (Biganzioli et al., 2018), proposed by the European Union. The software used for the study has been Simapro 9.0, with ecoinvent 3.5 as database for all the background data (Wernet et al., 2016). Processes included in this database are used for most of the steps included in this study, except for the process for UMG-Si production, which have been set from scratch. All these processes have been updated to the best of the authors' knowledge, recurring to up-to-date literature or data provided by industry, to account for the technical improvements that have been developed in the past years by the PV industry. The key parameters used for this study, according to the IEA guidelines, are shown in Table 1.

*Table 1. LCA Key Parameters*

| | |
|---:|:---|
| **PV technology** | Silicon feedstock: UMG and polysilicon<br>Multicrystalline Al-BSF cell technology |
| **Production location** | Processes:　　　　Europe<br>Electricity:　　　　Europe (Spanish mix)<br>　　　　　　　　　China (Chinese mix) |
| **Type of system** | Ground-mount, fixed-tilt |
| **Module efficiency** | UMG: 18.43%<br>Polysilicon: 18.55% |
| **Module degradation rate** | 0.4% |
| **PV lifetime** | 30 years |
| **BOS lifetime** | 15 years electrical gear<br>30 years structure |
| **Location of installation** | Tabernas (Almería, Spain) |
| **Annual irradiation** | 2160 kWh/m$^2$ year |
| **Inclination / Performance Ratio** | Optimal tilt angle 34º / 82,5% |
| **Time frame of data** | 2015-2020 |

### 2.1 Goal and Scope

The main goal of the study is to characterize the environmental impact associated to the use of solar grade silicon produced by a metallurgical direct route (UMG) by the Ferrosolar process for electricity production. The functional unit used in this study is 1 kWh of electricity produced by a ground mounted multicrystalline PV system with a nominal capacity of 100 MW. The selected functional units for each step included within the boundaries of the studied system can be found in the Supplementary Material (S1).

This "cradle to use" system includes the following items:

- Production of raw material: metallurgical grade silicon from silica sand
- Production of the solar grade silicon feedstock: Ferrosolar UMG silicon and polysilicon for solar applications
- Production of multicrystalline silicon ingots and wafers
- Al-BSF (Back Surface Field) multicrystalline solar cell manufacturing
- Multicrystalline PV module manufacturing
- Balance of system (BoS), ground mounting and electrical gear for a 100MW plant
- Installation, operation and maintenance of the PV system

In general terms, transport of the different materials between production steps has been not considered, with exception of PV modules transport to PV site location. Production sites do not have a defined location and may be very close to each other. Moreover, it has been shown before that transportation only accounts for a minor part of the final impact in electricity (Stamford and Azapagic, 2018). Packaging has been taken into account only for silicon wafers, due to their fragility, and PV modules that have to be protected in order to be transported to the PV site.

End-of-life (EoL) has been taken out of this study due to the lack of industrial scale data, as this PV systems have not still reached the minimum volume needed for economical reutilization and recycling processes that will be needed by the end of the next decade (Chowdhury et al., 2020). No difference should be found in recycling between polysilicon and UMG-Si as the recovery of silicon is expected to provide low grade silicon (similar to metallurgical grade) (Latunussa et al., 2016).

An uncertainty analysis was not performed in this study as no coherent sets of values where available for all the stages. Previous results indicate that the main factor for uncertainty is electrical consumption and the most affected stage is the silicon feedstock production (Huang et al., 2017).

Regarding the environmental impacts, all categories present in the EF method were calculated, but the focus was set on Climate Change. Additionally, Acidification terrestrial and freshwater (ATF) results are presented. To complement this result, cumulative energy demand (CED), total and non-renewable, was also assessed, together with the energy pay-back time (EPBT) for the considered scenarios. This parameter, expressed in years, was calculated as per equation (1) (Frischknecht et al., 2016):

$$EPBT = \frac{CED}{\frac{O_{e,yr}}{\eta_G}} \qquad (1)$$

Where:

*CED*: Total cumulative primary energy demand for the considered system [kWh].

$O_{e,year}$: Average electricity delivered by the considered system over the course of one year of its lifetime [kWh·year$^{-1}$].

$\eta_G$: Life-cycle energy efficiency of the grid mix into which the considered system is embedded (a value of 0.31 is assumed).

## 2.2 LCI data

All background inventory data is taken from ecoinvent 3.5 database. The selection criteria for processes in the database has been using Europe [RER] values when available, and Global [GLO] when no specific values for Europe where available. In some cases, only values for Rest of the World [RoW] where available. The full life cycle inventory data can be found in the Supplementary Material (S3). Data sources and assumptions are explained in the following sections for the selected foreground processes.

### 2.2.1 Electricity mix

All calculations have been carried out for both UMG and polysilicon feedstocks considering two different electricity mixes: Spanish (ES) and Chinese (CN) (Supplementary Material S2). Spanish mix (REE, 2019) has been selected because Ferrosolar process is meant to be carried out in the facilities located at Puertollano, Spain (Forniés et al., 2019), and can be considered among lower carbon intensity mixes. Chinese mix has been selected for comparison as China is currently the major PV devices manufacturer, reaching a share of the market of over 70% in all stages of the value chain (Philipps and Warmuth, 2020). Chinese mix is high-carbon intensity mix, with a high contribution of fossil fuels-based electricity production (BP, 2019). Electricity production for low, medium and high voltage supply, for both mixes, has been included in the LCA.

No other difference has been made between the scenarios calculated for ES and CN electricity mix, as it is not in the scope of this work to evaluate the impact of the manufacturing locations of the different stages and would introduce additional variability in the results.

### 2.2.2 Raw Materials: Silica Sand and Metallurgical grade silicon

The metallurgical production of silicon (MG-Si) implies the reduction of quartz with carbon in electric arc furnaces to obtain the silicon. Some wood chips are normally added to the reacting mix too, in order to get permeability in the mass and promote reaction between gaseous intermediates present in the furnace, as $SiO_{(g)}$ and $CO_{(g)}$ (Schei et al., 1998). Depending on the type of coal used in the process, the amount of $CO_2$ emissions coming from it can be very different.

Extractive mining of quartz main inputs are petrol based fuels and electricity, both of which depend significatively on the mine characteristics and the processes employed for product beneficiation (Grbeš, 2015). The data available in ecoinvent database has been used in this work, as no more meaningful values for this study could be found.

Metallurgical grade silicon production is a mature process from a technical point of view. Values from ecoinvent have been used for inventory, with some minor changes. The values presented in this database are representative of the Norwegian producer (Elkem). Electricity consumption diverges from some values for Chinese production that can be found in literature (Huang et al., 2017) but are in good agreement with experts' criteria. The manufacturing infrastructure has been changed from "silicone plant" to "electric arc furnace" as silicon production belongs to the metallurgic sector and not to the chemical one, as silicones do.

### 2.2.3   Silicon feedstock: Polysilicon

Solar grade silicon used by industry as silicon source for crystalline silicon PV devices manufacturing at the present time is produced mainly by a closed-loop Siemens process, in which trichlorosilane Siemens CVD deposition technology is combined with hydrochlorination of silicon tetrachloride for recovery of vent gases. This is the technology used by leading SoG-Si producers in China (Xie et al., 2018). Traditional western manufacturers as Wacker Chemie AG (Germany) or Hemlock Semiconductor (US), which account for around one quarter of the worldwide total production, use other technological schemes that rely on a high integration of chlorine use and production of chlorosilanes for other industries. Moreover, Chinese manufacturers are increasing each year their installed capacity. Polysilicon production technology has seen a very intense reduction of energy consumption and cash costs over the past decade (Chunduri, 2017; Woodhouse et al., 2019), although not updated values for input consumption can be found in open literature. The inventory data for closed loop hydrochlorination process adjusted to meet current material and energy consumption that has been used in this study can be found in the Supplementary Material (S3). Estimated quantities of nitrogen and graphite have been also included in the inventory, as they are known to be used by all polysilicon producers.

### 2.2.4   Silicon feedstock: UMG-Silicon

Ferrosolar's UMG silicon for solar applications has a boron concentration bellow 0.2 ppmw and a phosphorous concentration that can be tailored between 0.1 and 0.3 ppmw. Metals concentration is below 0.5 ppmw (including Fe, Al, transition, alkaline and alkaline earth elements). These specifications are suited for multicrystalline silicon applications.

The UMG production process description can be found in previous works (Forniés et al., 2019). For inventory purposes the process has been divided in its main steps: slagging, vacuum refining and directional solidification. Two additional processes have been modelled: an inert melting step, whose purpose is to recycle non pure Si material from the final steps of the process to reduce material consumption, and a final step in which the final formulation of the product is obtained. A simplified block diagram for the process is shown in Figure 1, illustrating the main inputs and output of the process. Further details cannot be disclosed in this work due to confidentiality reasons.

Mass allocation has been applied to by-products, as avoidance or economical criterions are difficult to formulate and can be considered a source of arbitrariness.

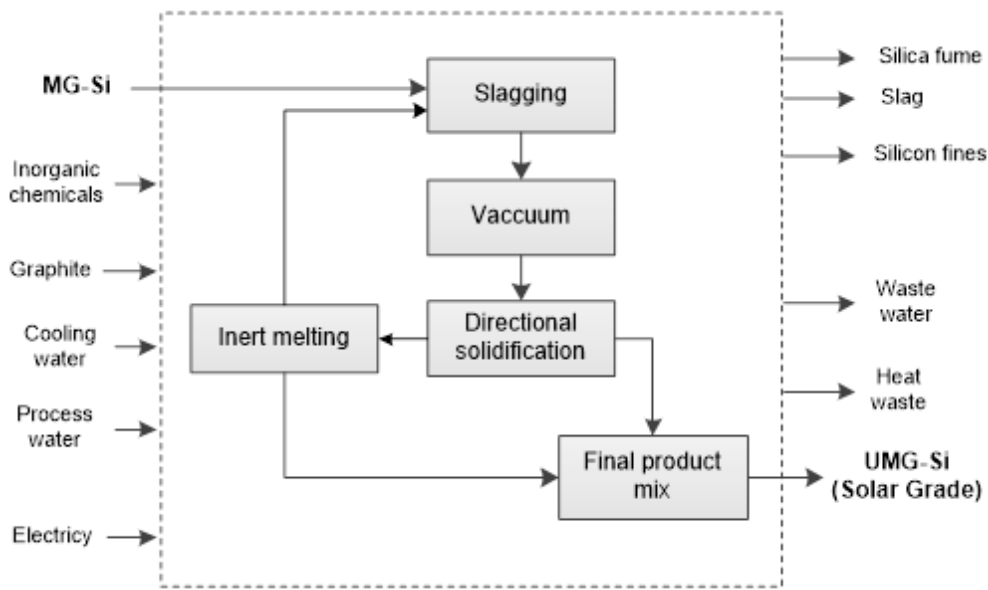

*Figure 1. UMG-Silicon production diagram: main inputs and outputs*

### 2.2.5 Multicrystalline silicon ingot and wafer manufacturing

As it has been pointed out, polysilicon, silicon ingot and wafer production, have seen an important improvement in their yield and energy consumption, driven by optimization of processes and equipment. In this study only the multicrystalline silicon process is analysed because UMG Silicon application without blending is limited to this technology, due to its higher level of impurities when compared with high quality polysilicon. High performance multicrystalline casting (HP-mc) (Lan, 2019) is the standard method used nowadays by industry and can be industrially applied to UMG silicon (Buchovska et al., 2017).

Regarding the wafering process, the most important development, diamond wire sawing (DWS) has been included in the LCI. The implementation of DWS has been very disruptive, becoming the only technique used for multicrystalline silicon wafering by 2019 (VDMA, 2019). This cutting technology relies on a wire that contains, embedded in its surface, synthetic diamond particles (Diamond-Like Carbon, DLC). Therefore, in order to cut the silicon ingot, it does not require an abrasive suspension. In the case of traditional cutting by Multiwire Slurry Saw (MWSS), the cut is made using a metal thread that presses an abrasive suspension, called slurry, on the ingot, which contains polyethylene glycol (PEG) and silicon carbide particles. One of the main advantages of DWS is the reduction of kerf loss, since the groove left by the diamond wire is about 80 microns compared to the 120-200 microns of the traditional slurry cut, leading to a productivity of 45 wafers per kg of ingot compared to 60 wafers per kg of MWSS (Rentsch et al., 2018). Multicrystalline silicon is less suited for this technique than monocrystalline silicon but this drawback has been overcome by the improvement of the wet chemical processes needed for texturing and saw damage removal (Shetty et al., 2020).

For the LCI formulation of the wafering process main assumptions include a silicon consumption of 0.665 kg per $1m^2$ of wafer, for 156.75 x 156.75 mm wafers. The yield of the process, in terms of silicon mass, is over 60%, much higher than the 40% achieved by MWSS technique.

### 2.2.6 Al-BSF Cell and PV Module manufacturing

Standard Aluminium Back Surface Field (Al-BSF) technology for PV solar cells manufacturing has been considered. Details regarding said process can be found in a previous work (Fornies, Energies 2019). It is assumed that the needed amount of wafers per cell is 1.02 $m^2$ wafer/ $m^2$ cell. Module manufacturing has been developed intensively over the past few years in a pursue of a lower cost-higher efficiency devices. The characteristic parameters of the PV modules used for this study are shown in Table 2. A loss from broken cells of 1.67% has been considered. The LCI tables for both processes are available in the Supplementary Material (S3).

*Table 2. PV Module characteristic parameters*

|  |  | UMG | polysilicon |
|---|---|---|---|
| **Maximum Power at STC (\*)** | $W_p$ | 325.88 | 328.08 |
| **Module size** | $m^2$ | 1.938396 | |
| **Mass** | kg | 22 ± 3% | |
| **Glass thickness** | mm | 3.2 | |
| **Cell number** |  | 72 | |
| **Solar cell efficiency** | % | 18,52 | 18,64 |
| **Solar cell size** | mm | 156,75 | |
| **Bus bar** |  | 5 | |
| **CTM (\*\*) losses** | % | 99.5 | |

(\*) Factory Standard Test Conditions": 1,000 W/m$^2$ solar irradiance, 1.5 AM, 25ºC

(\*\*) Cell-to-module

### 2.2.7 Balance of system, installation and plant operation

The BoS includes all the necessary equipment to install the PV modules in the desired configuration and transform the electricity to its final use. The following processes have been contemplated: site conditioning, mounting structure, electrical installation and inverters, PV panel transportation and installation.

Inventory for mounting structure and electrical installation has been calculated from current real data gathered from TINOSA Project, comprised of several plants, with a total installed power of 180MW. This PV site, located in Almería (Spain), is currently under construction by Aurinka PV International (Sánchez Molina, 2020).

The modelled PV plant has a total installed power of 100 MW$_{pk}$, a fixed structure and a set of PV skids comprised of central inverter, switchgear and transformer. Production values are also calculated for the ubication of the mentioned project. A reduction of 0.5% in the produced electricity due to self-consumption was set for EPBT calculations.

## 3 Results and discussion

Tables reporting the full results obtained for the life cycle assessment can be found in the Supplementary Material (S4) for all process stages studied in this work.

### 3.1 LCIA Results

#### 3.1.1 Electricity mix

The results for electricity mixes used in this study are first discussed, in order to support the analysis of the results obtained for the rest of the analysed processes, some of which have large electricity consumptions. Results can be found in the Supplementary Material (S5). The impacts considered for Spanish (ES) and Chinese (CN) electricity mixes present very large differences, about 3 and 10 times higher for the latter for CC and ATF categories respectively. This essentially is the result of the different weight of fossil-based electricity production technologies. Coal (hard coal and lignite) is responsible for more than 90% of CC emissions and ATF, in the case of CN mix, whereas natural gas accounts for around 60% of CC for ES mix, that adds up to the contribution of coal (around 20%) and oil (5%). Nuclear and renewables (60% of the mix) do not add significative CC emissions in this case.

On the other hand, the coal contribution to ATF in ES mix goes over 75%, despite its relatively low share in the mix (about 5%), thus giving a low absolute total value when compare to CN mix. CED comparison shows much less differences, being 15% higher for CN, relying mainly on coal instead of natural gas combined cycle and nuclear, as non-renewable generation technologies. CC emissions calculated for Spanish mix, 0.201 kgCO$_{2eq}$ / kWh$_e$, are very similar to those published by the responsible official entity (European Environment Agency, 2020), and can be considered among low intensity mixes, whereas Chinese mix, with 0.591 kgCO$_{2eq}$ / kWh$_e$, would belong to high carbon intensity mix group.

#### 3.1.2 PV Electricity Production

The results for PV electricity production in the assessed scenarios are presented in Figure 2. In the case UMG-ES, the value obtained for Climate Change (CC) category is 12.10 gCO$_{2eq}$ / kWh$_e$ and for Acidification (terrestrial and freshwater) (ATF) is $9.25 \cdot 10^{-5}$ molH$^+_{eq}$/ kWh$_e$, according to the definitions in the EF method. The result for Total Cumulative Energy Demand (CED) is $5.67 \cdot 10^{-2}$ kWh/ kWh$_e$.

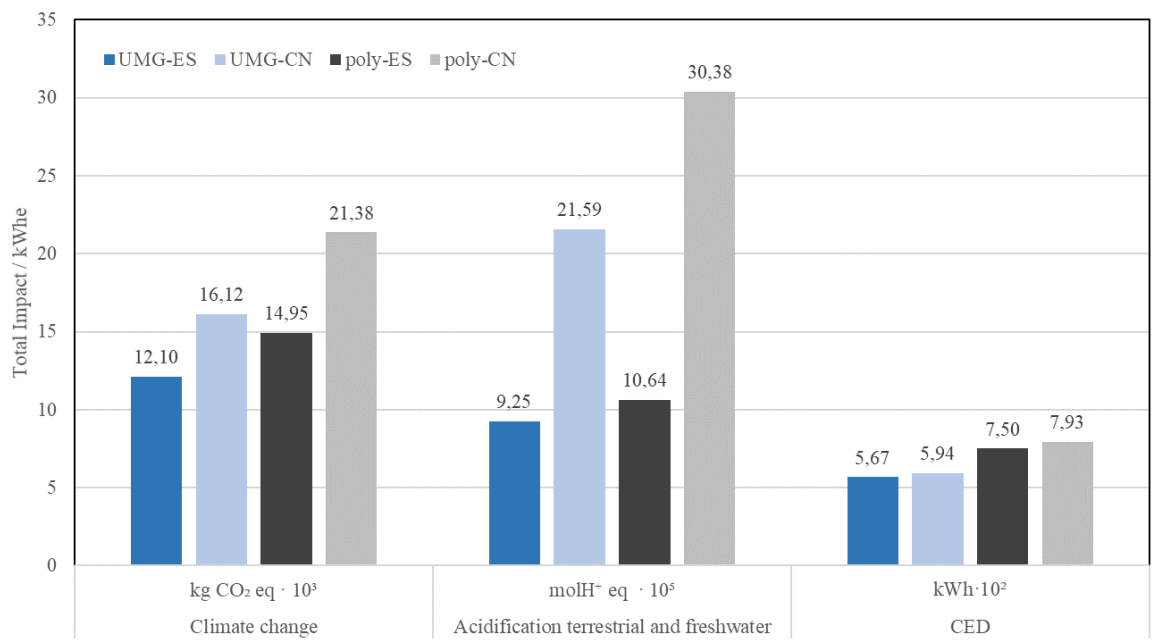

*Figure 2. Main impacts per kWh of produced electricity for the defined scenarios (100MW@2100W/m²year)*

Predictably, due to the higher carbon intensity of its energy mix, UMG-CN CC category result is 33% higher than UMG-ES. As explained in the previous section, the only difference considered in the LCI between these scenarios is the electricity mix. ATF category behaves in a similar manner, although in this case UMG-CN is 133% higher than UMG-ES. This is due to the important share of hard coal in Chinese mix, in contrast to the relatively low share in the Spanish mix (see Supplementary Material S5). On the other hand, when the CED is compared the differences are not as significant, achieving UMG-ES scenario a result less than 5% lower than UMG-CN.

The same trend is observed when poly-ES and poly-CN scenarios are compared, intensified by the higher energy demand, mostly electricity, of these scenarios when compared to the former. This results in a difference of 43% in CC, 185% in ATF and 6% in CED categories, between poly-ES and poly-CN scenarios.

The comparison of UMG-ES/poly-ES and UMG-CN/poly-CN scenarios indicate that in the cases where UMG-Si was used as feedstock, electricity production impacts are noteworthy lower, although the differences where not as remarkable as in the previous comparison for ATP, being more significant in the case of CED. The differences between the pairs UMG-ES/poly-ES and UMG-CN/poly-CN are 24% and 15% for CC and 33% and 41% for ATF respectively. The observed results can be attributed to the higher electricity and heat demand characteristic of polysilicon manufacture, that will be shown in detail in next section.

Climate change calculated emissions vary between 12.1 and 21.4 $gCO_{2eq}$/ $kWh_e$ for UMG-ES and poly-CN scenarios. Comparing with previous studies that can be found in literature shows that the results are somewhat lower than those published before. It is important to note that these comparisons are not straight forward, as parameters that characterize each case study may differ, having some of them very important influence on the overall result, as for example the efficiency of the solar cell, and therefore of the PV module, the irradiation of the location selected or size of the installation, which may vary from less than 10 $kW_{pk}$ for domestic installations to hundreds of MWp for utility-scale PV parks. It is also common in literature to find studies in which the type of silicon used is a mixture of both mono and multicrystalline materials. Moreover, different systems boundaries are defined, and various LCA assessment methods are used, leading to different results.

In order to put the obtained outcomes into context, a set of case studies from literature have been selected and their conditions and results gathered in the Supplementary Material (S6). The criteria for selection have been that the systems considered resembled the present study as much as possible. We are aware that this evaluation is problematic, for example for climate change emissions, as assessment methods have different definitions for impact categories, but nonetheless it provides useful information for discussion. This was in fact the main reason to evaluate four different scenarios in this work, ensuring that the results would be comparable to each other. Some very complete works (Hong et al., 2016; Huang et al., 2017; Xie et al., 2018) have not been included as they have applied different LCA methods, as they provide non comparable results to ours.

In the case of UMG feedstock, literature values are scarce. Elkem Solar, back in 2012 (de Wild-Scholten and Gløckner, 2012), reported values for UMG electricity production GWP of about 29 $gCO_{2eq}$ / $kWh_e$. It is to note that Elkem used for their calculations of the UMG silicon feedstock impact the Norwegian electricity mix, which has an especially low carbon intensity as it relies mainly on hydropower. In a more recent work (Yu et al., 2017), a value of 20 $gCO_{2eq}$ / $kWh_e$ was reported which is in line with the results of the present work, 16.1 $gCO_{2eq}$ / $kWh_e$ when Chinese mix is considered. Yu et.al. considered a process for UMG silicon feedstock production that differs from the Ferrosolar process, including two steps of directional solidification and an electron beam melting step in between. Both processes have in common a low energy consumption when compared to the conventional modified Siemens process for production of solar grade silicon. Additionally, the conditions used in this study are more favourable: apart from better irradiation and bigger size of the installation for the selected site, the used of higher power modules with better performance along its lifetime have an important influence on the results.

Regarding polysilicon performance, more information is available in literature for a validation of the results. A recent study (Raugei et al., 2020) discussing the impact of the inclusion of batteries to the PV systems, come up with a value for GWP of 27 $gCO_{2eq}$/ $kWh_e$ (37 $gCO_{2eq}$/ $kWh_e$ @1700 $kWh/m^2$·year). In their work they use a mixture multi and monocrystalline silicon feedstock with a distribution of 55%-65%, of Chinese provenance, including a similar electricity mix to CN mix in the present work. In another study, dealing with the comparison of fixed and single axis tracking PV systems environmental impacts (Antonanzas et al., 2019) obtained a value for $CO_2$ emissions of 23.0 $gCO_{2eq}$/ $kWh_e$. These results are slightly higher than 21.4 $gCO_{2eq}$/ $kWh_e$. This may be attributed to the update of some important inputs of the life cycle inventory of the different stages that the PV value chain comprises, as older studies (Fu et al., 2015) show higher scores than more recent ones.

*3.1.3 PV Value Chain*

In order to gain insight of the reasons behind the obtained overall result for PV electricity production, an analysis of the contribution of the different production stages is carried out. The share of each process for CC, ATF and CED is shown in Figure 3, Figure 4 and Figure 5, respectively. A full table can be found in the Supplementary Material (S7). The result of each stage excludes the contribution of the previous one: ingot production contribution does not include the silicon feedstock, PV cell fabrication does not include the contribution of the multicrystalline wafer production, and so on.

The absolute contribution of the BoS and Installation to the overall result are the same in all the scenarios, as only one location for the PV site has been considered. It is to be noted that the inputs for the inventory have been obtained from a specific project (TINOSA plant, in Almería (Spain)) and are therefore conditioned by the characteristics of the site: i.e., wind velocity, ground slope profiles… On the other hand, the values used are representative of the state-of-the art regarding BoS of large size ground installations. The calculated need of structural steel is 42 kg per $kW_{pk}$ installed, which is of the same order as 31 kg/$kW_{pk}$ (Mason et al., 2006) and very similar to the 38 kg/ $kW_{pk}$ calculated recently by Antonanzas et.al.

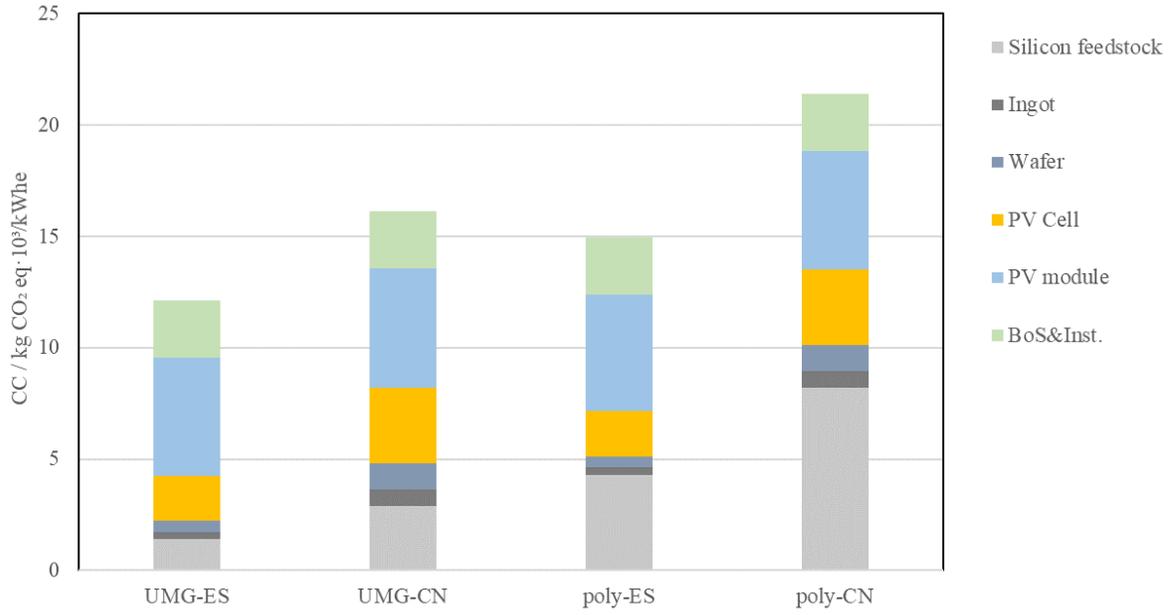

*Figure 3. Climate change contributions of different stages of the production of PV systems per kWh of produced electricity*

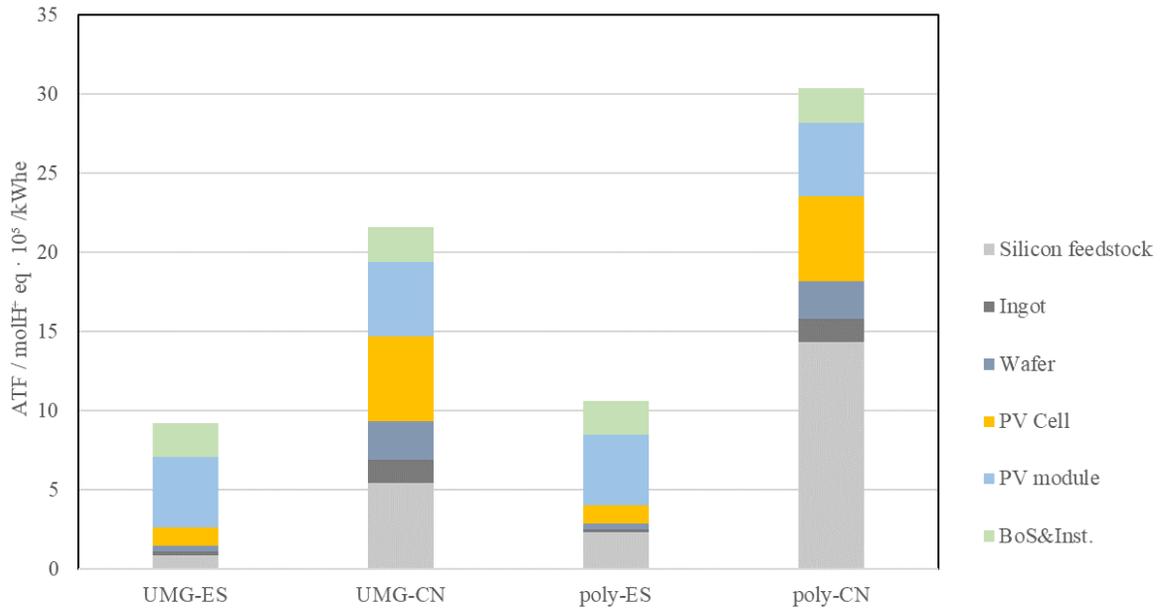

*Figure 4. Acidification terrestrial and freshwater: Contributions of different stages of the production of PV systems per kWh of produced electricity*

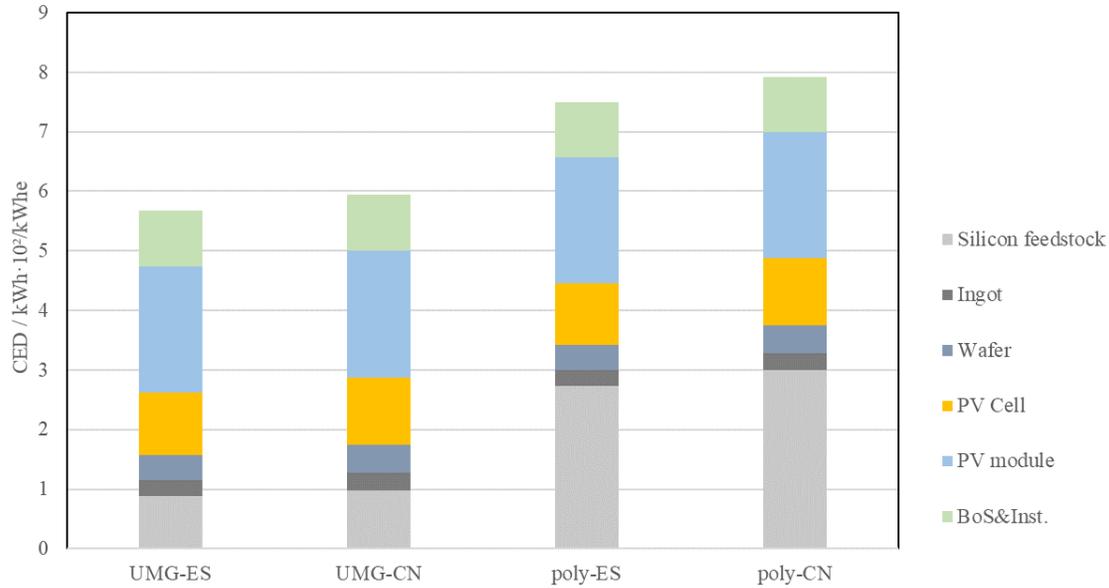

*Figure 5. Cumulative Energy Demand: Contributions of different stages of the production of PV systems per kWh of produced electricity*

The calculated climate change emissions of BoS and Installation account for around 130 kg $CO_{2eq}$ / kWp of installed power in the PV site, in contrast for example to the 494 kg $CO_{2eq}$ / kWp coming from the PV Modules in UMG-ES or the 974 kg $CO_{2eq}$ / kW$_{pk}$ in poly-CN scenarios, best and worst cases of the set. The obtained values are similar to recent results for ground mounted fixed structures: 143.7 kg$C_{2eq}$ /kW (Antonanzas et al., 2019) and much lower than older studies: about 300 kg$CO_{2eq}$ about/kW (de Wild-Scholten, 2013; Leccisi et al., 2016). This reduction is the result of several positive effects combined, the increase of the ratio of power and weight of the PV modules power over the years, resulting in less demanding need of structural components and electrical gear per installed kW$_{pk}$ and the improvements in the components fabrication itself, avoiding the use of aluminium and concrete in fixed installations and cooper for electrical cables (substituted by aluminium where possible) and optimizing the needed amount of materials. This has been pointed out before by Antonanzas et.al. as many works rely on ecoinvent data (Wernet et al., 2016), that must be corrected according to the power of the PV modules. As it is shown in Figure 6,the PV panel itself account for about 80% of the impact, being the other major contributors the mounting system, the electrical gear and the inverter.

Regarding PV module manufacturing, impacts are also very similar in all the scenarios for both CC and ATF categories (see Table 3). On the one hand, the energy consumption is low when compared to the rest of process steps. On the other hand, the difference in calculated module power at STC conditions for UMG and polysilicon modules is about 0,67%, according to the efficiency stablished in a previous work (Forniés et al., 2019) , which was 18,515% for UMG based cells and 18,640% for standard polysilicon-based ones. This outcome on slightly higher values for UMG when compared to poly scenarios, around 1%, due to the decrease of efficiency. CED values are also mostly the same for all the scenarios. As can be seen in Figure 6, apart from the PV cell, the main contributors to CC and ATF are the aluminium frame, the backsheet and the solar glass, thus the impact is not affected either by the silicon feedstock or the electricity mix. EVA does also have an important contribution to CED, while cooper does to ATF.

Results for polysilicon are comparable to those found recently in published works. Luo et.al obtained a CC of 215.9. kg $CO_{2eq}$ /module (219.5 for PERC modules), applying Singapore electricity mix (0.4846 kg $CO_{2eq}$ /kWh) (Luo et al., 2018). Corresponding values in this study are 210.4 and 228.7 kg $CO_{2eq}$ /module, for ES and CN mix respectively. Recently NREC (Korea's New and Renewable Energy Centre) classified Chinese PV modules in a category with over 830 kg $CO_{2eq}$ /kW$_p$ (Stoker, 2020). The result is highly influenced by the efficiency of the PV cell, as the same quantity of materials and energy are used to obtain modules with varying nominal power. Sooner or later, regulations considering the carbon footprint of PV modules are likely to be stablished in more countries, other than France and South Korea, in particular in the European Union, making UMG-Si PV modules more attractive to the end users, when compared to polysilicon.

Although this work does not account for PERC processing, the CC emissions for UMG modules can be estimated from Luo et.al. results. In their work they found a difference between PERC and BSF of 1.67%. In our previous work, a mean efficiency of 20.13% for UMG black silicon PERC cells was obtained, which would correspond to a 354.3 W module. Assuming no other differences, the CC emission for this module result in 478 kg$CO_{2eq}$/kW, showing a reduction of 3%.

Regarding the absolute total results for ingot, wafer and cell manufacture steps, consumption of electrical power is relevant, leading to big differences of the impact calculated for scenarios with different electricity mixes. As no distinction is considered in the processing of UMG and polysilicon for ingot, wafer and cell production, variations are caused solely by the energy mix used for calculations, resulting in very large differences between ES and CN scenarios, about 85% higher values for the later in the case of CC. The same behaviour is observed in the ATF category results, intensified by the big difference between ES and CN mixes for this category, leading to values over 4.5 times higher for these processes when Chinese mix is employed.

The critical role of electricity in the impact of these processes is clearly shown in Figure 6 (c, d and e). In the case of PV cells and wafers it accounts for about 15% of the total impact and a little less for silicon multicrystalline ingots for both CC and ATF categories.

The trends observed for CC and ATF are not found in CED results for the different scenarios, as this parameter is only slightly affected by the electricity mix. As could be expected, a big discrepancy is found between UMG and poly scenarios, whereas small variations are observed when the Spanish mix is compared to the Chinese one. The use of UMG instead of polysilicon leads to a reduction of 33% in the CED per $kWh_e$. This is of much importance in the calculation of the energy payback time of the different technologies. For UMG scenarios the main contributor to CED is the PV module fabrication step, followed by PV cell production, while for poly scenarios it is the silicon feedstock the largest contributor.

The obtained results are difficult to compare to those than can be found in open literature, because even the more recent studies do not include the use of DWS, and still rely on data from MWSS processes, essentially taken from databases or previous literature(Fan et al., 2020; Raugei et al., 2020). The wafering step contribution to the total impact is much less with DWS as it improves the yield very much while reducing the need of other materials such as silicon carbide and PEG.

Regarding the impact of the silicon feedstock Fan et.al. studied four scenarios including both materials, for multi and monocrystalline wafers. Their results show that ratios between multicrystalline UMG and poly are 0.235 for PED (Primary Energy Demand) and 0.267 for $CO_2$ emissions. In the present work the calculated ratios are 0.432 and 0.456, respectively, higher because of the lower impact of the wafering process.

Multicrystalline ingot casting, apart from the mentioned contribution of electrical power, has a significant one by argon. This consumption (0.75 kg/ kg ingot) was found out to be significantly lower than some literature values (1.92 kg/kg ingot (Fu et al., 2015)) but also higher than others (0.25 kg/kg ingot (Frischknecht et al., 2015)). The difference with the later might be attributed to the introduction of the high performance multi casting.

*Table 3. Contribution of the different process step to Climate Change (per $m^2$ and kW of nominal PV module power)*

|  | Silicon | Ingot | Wafer | PV Cell | PV module | Total |
|---|---|---|---|---|---|---|
| **Climate change ($kgCO_{2eq}/m^2$)** | | | | | | |
| **UMG ES** | 12,1 | 2,9 | 4,4 | 17,6 | 45,9 | 83,0 |
| **UMG CN** | 25,0 | 6,5 | 10,2 | 29,6 | 46,7 | 117,9 |
| **Poly ES** | 37,6 | 2,9 | 4,4 | 17,6 | 45,9 | 108,4 |
| **Poly CN** | 71,8 | 6,5 | 10,2 | 29,6 | 46,7 | 164,7 |
| **Climate change ($kgCO_{2eq}/kW@STC$)** | | | | | | |
| **UMG ES** | 72 | 17 | 26 | 105 | 273 | 493 |
| **UMG CN** | 149 | 38 | 61 | 176 | 278 | 701 |
| **Poly ES** | 222 | 17 | 26 | 104 | 271 | 641 |
| **Poly CN** | 424 | 38 | 60 | 175 | 276 | 973 |

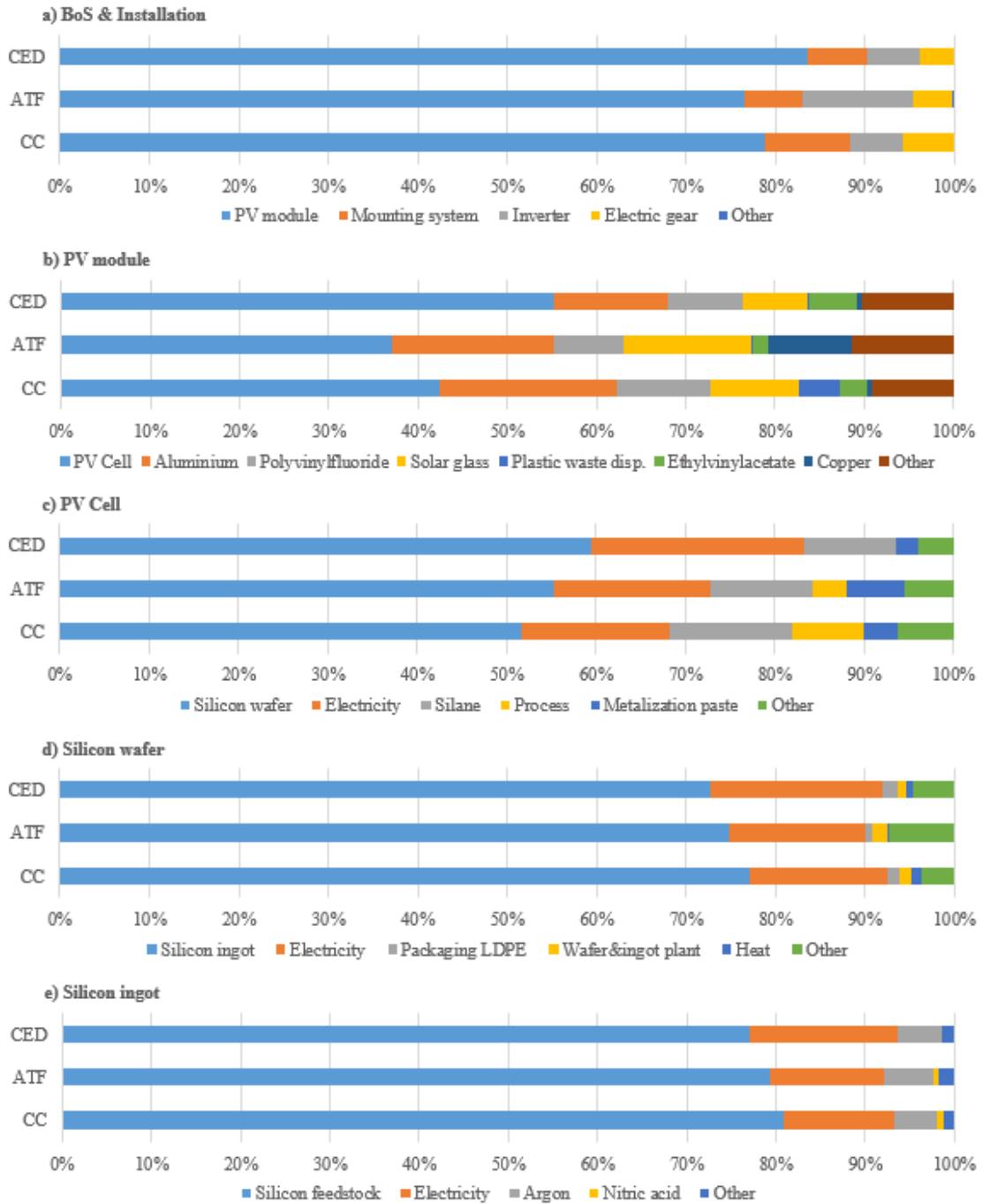

*Figure 6. Detailed individual contribution breakdown for the different steps of the PV value chain (UMG-ES scenario)*

*3.1.4 Silicon feedstock: UMG vs. polysilicon*

It has already been revealed that UMG based PV electricity has less carbon footprint and a smaller demand of energy than polysilicon, but in order to explain the differences a detailed comparative analysis of both materials is carried out. The full results of the assessment for UMG-Si and polysilicon can be found in the Supplementary Material (S4), both for ES and CN electricity mixes. CC, ATF and CED results per kg of SoG-Si produced are shown in Figure 7. As expected, polysilicon scores higher in every category, as it is reported in previous studies (Yu et al., 2017). UMG-ES calculated impacts are the lowest of the four scenarios. These values are 17.9 kg $CO_{2eq}$/kg for CC and $11.2 \cdot 10^{-2}$ $molH^+_{eq}$/kg for ATF category, according to the definitions in the EF method. The result for CED is 113 kWh/kg. On the other side, poly-CN has the highest impacts: CC 106.2 $kgCO_{2eq}$/kg, ATF $185.3 \cdot 10^{-2}$ $molH^+_{eq}$/kg and CED is 387 kWh/kg.

The trends observed for ATF and CED are very similar to those found for PV electricity. For ATF, we find higher value in CN mix scenarios, being the contribution of this coal intensive generation more important than the process for SoG-Si production itself. That means that UMG-Si obtained applying CN mix would have more impact in terms of ATF than the polysilicon obtained with a mix including less share of coal, $69.8 \cdot 10^{-2}$ against 29.9 $molH^+_{eq}$/kg. On the contrary, it is the SoG-Si production process the determining factor for CED results, those obtained for polysilicon are about 3 times higher than for UMG. This is related to a higher electrical power consumption (85 kWh/kg for polysilicon and 26.5 kWh/kg for Ferrosolar UMG considered process) and the use of heat as input in polysilicon production (180 MJ/kg). The electrical mix is solely responsible for the differences found between the results obtained for each material.

Regarding Climate Change (Figure 3), the discussion of the results is not straight forward although the influence of the electricity mix is clear as the CC for CN mix double those for ES mix, both for UMG and polysilicon. The contribution of the main inventory inputs for UMG and polysilicon production processes is shown in Figure 8. Metallurgical grade silicon (MG-Si), electricity and heat in the case of polysilicon are the most important contributors to this category.

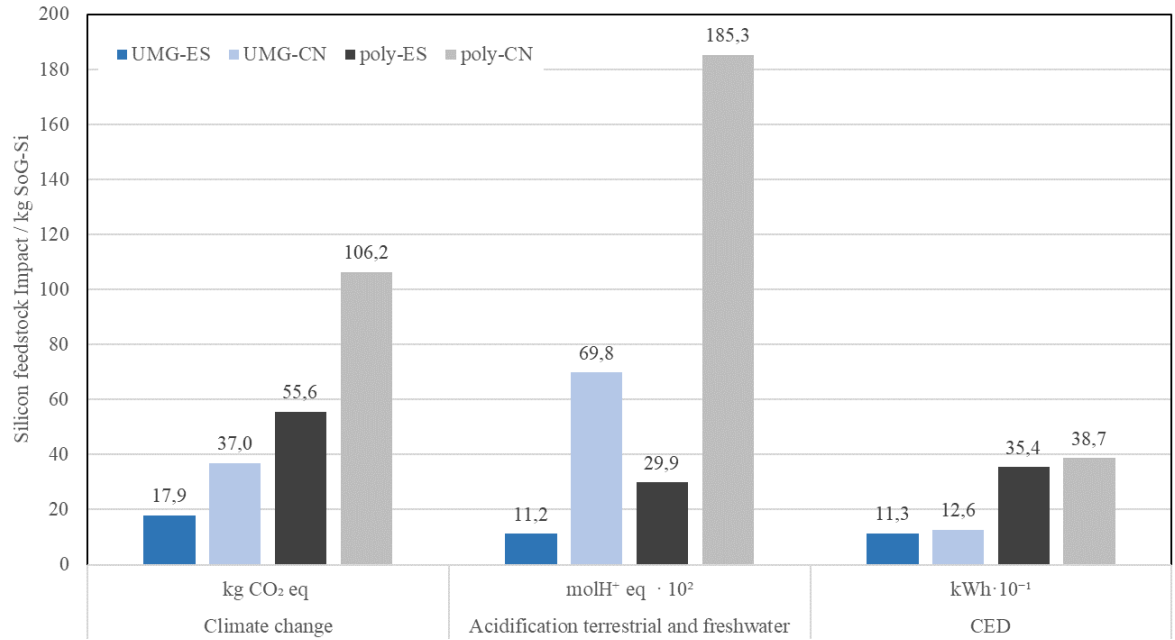

*Figure 7. Main impacts per kg of SoG-Si for UMG and polysilicon with ES and CN electricity mixes*

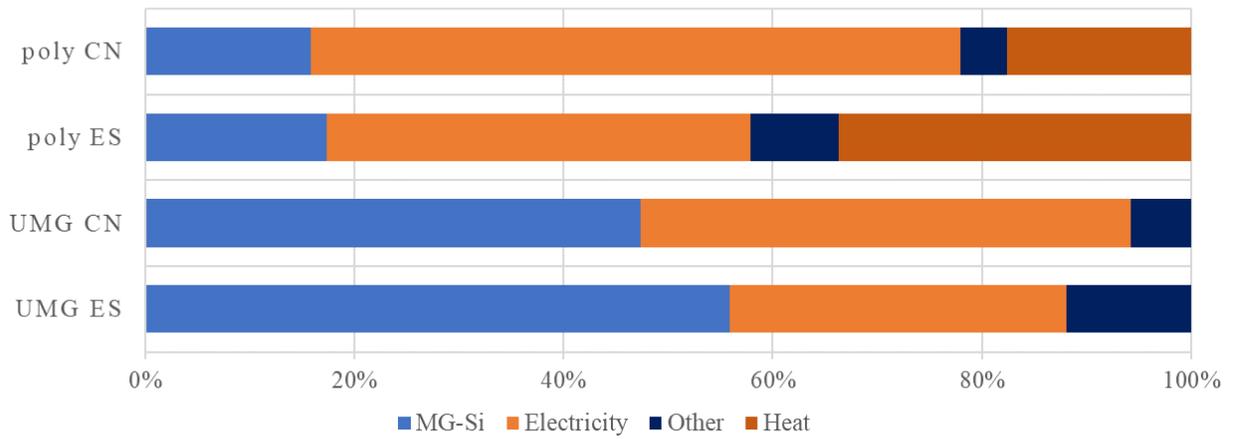

*Figure 8. Climate change distribution of main inputs for UMG and polysilicon*

Metallurgical grade silicon is the starting raw material for both, polysilicon and UMG obtention, although these processes differ on the amount of raw material needed to obtain a kg of usable product, that is higher in the case of UMG Ferrosolar process. This quantity may vary as it depends on the specific process studied. In the work of Yu.et.al. the ratio MG-Si to SoG-Si is 1.23 (Yu et al., 2017) for UMG, smaller and very close to the ratio used in this work for polysilicon, 1.26 (Woodhouse et al., 2019). The latter, is reduced in contrast to other values found in literature, for example 1.39 (Huang et al., 2017; Xie et al., 2018) although is higher than the value included in ecoinvent database: 1.13.

As this ratio has an important influence in CC, especially for UMG-Si, the breakdown of the contribution of the main inputs in the production of MG-Si has been calculated. The absolute values MG-Si are 7.65 and 13.29 kg $CO_{2eq}$/kg for ES and CN mixes, respectively and the two main contributions are electrical power (38% and 64% for ES and CN mixes), and process emissions (48% and 26%).

Process emissions are inherent to the oxidation of the carbonous material that is used for the reduction of silica. They may vary, depending on the nature of the coal mix used and have been studied before (Monsen et al., 1998) being possible the adjust the ratios between fossil and renewable carbon sources. On the other hand, electricity consumption is a key contributor to climate change emissions of MG-Si. As shown in Figure 8, about half of the CC impact in the case of UMG-Si comes from the MG-Si, so any improvement in its production process would lead to even lower emissions when compared to polysilicon, for which MG-Si CC contribution accounts only for around 15%.

Regarding energy consumption of SoG-Si production processes, it is important to note that polysilicon production has a high demand of heat (as steam) apart from electrical power. A heat input of 180 MJ and a electricity consumption of 85 kW/kg of polysilicon have been considered for this work, similar to those found in literature and ecoinvent database. If these values are further reduced, assuming 65 kWh and 90 MJ / kg of polysilicon, CC results would be 40.9 and 81.3 $kgCO_{2eq}$ / kg of polysilicon for ES and CN mixes respectively, still far from UMG results (17.9 and 37.0 $kgCO_{2eq}$ / kg of UMG).

The contribution to CC of the different steps used to model UMG production by the Ferrosolar process can be found in the Supplementary Material (S8). The share of the MG-Si used as starting material is for both (ES and CN) mixes the most important, as shown also in Figure 8. The slagging step, meant for boron removal, is also responsible for an essential part of the total CC impact, followed by vacuum refining, whose objective is phosphorous removal, and directional solidification, that serves as final purification of the UMG. All these steps have in common that they are electric power consumers, being the impacts very affected by the electricity mixed applied in the considered scenarios.

The comparison of the obtained results with published values is not easy as this kind of studies are few and aggregated impacts are provided. Recent work by Yu et.al does not provide any detail result for UMG, although they infer from PV production that this material is advantageous in terms of environmental impact (Yu et al., 2017). Values of 15 and 31 kgCO$_{2eq}$/kg UMG, for Norway (12 gCO$_{2eq}$/kWh) and UCTE (Europe) (530 gCO$_{2eq}$/kWh) electricity mixes were published for the Elkem Solar process (de Wild-Scholten and Gløckner, 2008) and updated afterwards to 10 kgCO$_{2eq}$/kg UMG (de Wild-Scholten and Gløckner, 2012), also with Norwegian electricity mix. Elkem Solar process is different than the one assessed in this work, including a leaching step instead of a vacuum refining of the intermediate product obtained after a slagging step. In the first work a detailed analysis of different configurations of the Siemens process was carried out, obtaining around 57 kgCO$_{2eq}$ / kg polysilicon obtained by a route named "TCS with dirty STC recycling" which corresponds to the modified Siemens process used in the present work, being electricity consumption (UCTE mix) responsible for about 80% of the total impact. In their more recent work, they indicate that the CC emissions of polysilicon, according to ecoinvent 2.2 database, are around 40 kgCO$_{2eq.}$ / kg polysilicon. The inventory included in ecoinvent 3 considers a total electricity consumption of 110 kW / kg polysilicon, of which 65 kWh/ kg are from hydroelectric power, having thus very small effect on the calculated CC impact.

3.2 <u>System Energy Payback Time</u>

The results of the assessment for Cumulative Energy Demand have already been presented for each stage of the PV value chain in the previous section, however additional detail is shown in Supplementary Material (S9) for further clarification. The part of the background processes (taken from ecoinvent database, not part of the modelled system) in the CED is found to be 8.1% of the total value for UMG scenarios and 6.4% for polysilicon ones. The background energy comes from the material used in the different stages, in particular from the components of the PV module: aluminium frame, solar glass, encapsulant and backsheet, which eventually account for over 10 kg /m$^2$ of PV module, or which is the same, over 90% of the total mass of each module.

EPBT for each scenario is calculated from these values and a mean annual electricity production of 1.82·10$^8$ kWh$_e$ /year for the 100 MW installed power PV site. 0.5% is subtracted in account for maintenance (9.12·10$^5$ kWh$_e$ /year). A total EPBT of 0.52, 0.55, 0.69 and 0.72 years are obtained for UMG ES and CN and poly ES and CN, respectively. The values of the nr-EPBT (non-renewable) and the r-EPBT (renewable) are shown in Figure 9. UMG based PV electricity has a less than 6 months of nr-EPBT, whereas polysilicon based are about 30% higher.

These results are significantly better than other published values for highly irradiated PV sites, both for UMG, as expected, but for polysilicon also. Similar values can be found in the work of Antonanzas et.al. for fixed CN manufacturing polysilicon PV site with a 2327 kWh/m$^2$·year, indicating a value of 0.65 years.

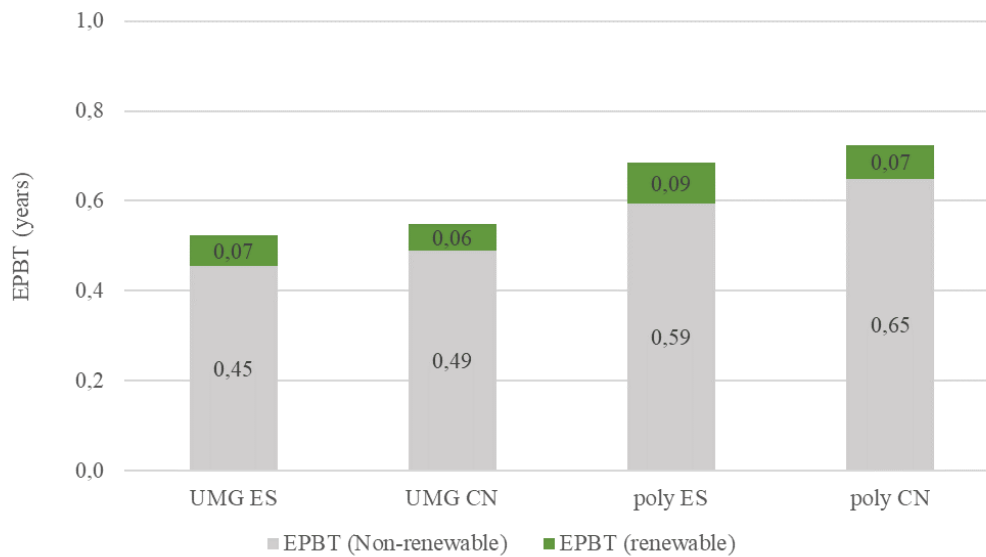

*Figure 9. EPBT (non-renewable and renewable) for the considered scenarios*

## 4 Conclusions

A comprehensive comparative life cycle analysis of UMG and polysilicon based PV electricity generation was performed "cradle to use". Both silicon feedstocks have been evaluated employing a low and a high carbon intensity electricity mix. The contribution of stages and components to environmental impacts was calculated by EF method. Cumulative energy demand was included in the analysis. The LCI where revised and updated to the best of the authors' knowledge, matching the production flow used in a previous work, in which the mean efficiency of an industrially manufactured UMG BSF solar cell achieved was over 18% efficiency, very close to that obtained for polysilicon. A large scale, 100MW, ground fixed PV site was considered, employing data a current project being developed in Almería (Spain) by Aurinka PV. This location is among the best in Europe for PV solar, as its annual in-plane irradiation is over 2000 kWh/m$^2$.

As was expected, the electricity mix plays a very important role in most impact categories, as the processes of which the crystalline silicon PV chain if comprised of demand substantial amounts of electricity, especially the silicon feedstock production. Acidification (ATF) impacts were specially affected by the electricity mix, as the coal intensive Chinese mix was compared to much lower impact Spanish mix.

The full PV value chain stages contribution was analysed, showing that when UMG is used as feedstock, its contribution to climate change, acidification and cumulative energy demand was not the highest, as it happened when polysilicon is used as feedstock and is usually reported in literature. PV module followed by cell manufacturing had the highest shares. In general, the calculated impacts are lower than those presented in previous studies, also for polysilicon, due to the update, most frequently reducing the quantity of materials and energy employed, of the inventories of the different stages of PV manufacturing. This is especially important in the wafering process, for which state of the art DWS technology has been used, that reduces the amount of silicon losses when compared to slurry based traditional process.

Among the most important results of the study where the verification that the use of Ferrosolar's UMG, instead of conventional polysilicon, can lead to a reduction of over 20% for climate change emissions for ES mix (12 vs. 15 g$CO_{2eq}$/ kWh$_e$) and over 25% for EPBT of the electricity generated (0.52 vs. 0.68 years) during the expected lifetime of PV system, considered 30 years. Regarding the silicon feedstock itself, the reduction of CC impact obtained was 50% for ES mix (18 vs. 56 g$CO_{2eq}$/ kWh$_e$), being electricity and metallurgical grade silicon the main contributors.

Although the results obtained for UMG are encouraging, more advanced technology should be evaluated to find the real potential of this material. PERC and half cells have not been included in current LCA study and are part of the next steps. A more detailed study of the BoS, especially inverters, would also provide useful results, as it plays an important part of the total impacts, due to the reduction of the PV module fabrication impact.

## 5   Funding

The authors received no financial support for the research, authorship, and/or publication of this article.

**CRediT author statement**

**Laura Méndez**: Conceptualization, Methodology, Investigation, Formal analysis, Writing- Original draft preparation, Visualization. **Eduardo Forniés**: Conceptualization, Methodology, Investigation, Writing - Review & Editing. **Daniel Garrain**: Methodology, Formal analysis, Resources, Writing - Review & Editing. **Antonio Pérez Vázquez**: Investigation, Writing - Review & Editing. **Alejandro Souto**: Investigation, Writing - Review & Editing. **Timur Vlasenko**: Investigation, Writing - Review & Editing.

**Highlights**

- Life Cycle Assessment methodology is applied to UMG silicon based large-scale PV site electricity generation.
- A comprehensive review and analysis of the full PV value chain is undertaken.
- A double comparison is carried out: UMG and polysilicon as silicon feedstock and two electricity mixes with low and high carbon intensity.
- Key factors that contributed the overall environmental burden are identified.
- Climate change emission of UMG PV electricity are 20% less than for polysilicon and energy payback time is decreased in 25%.

# Supplementary Material

## S1 – Functional Units

*Table 1. Functional Units for LCA foreground processes*

| Process Step | Functional Unit |
|---|---|
| **Metallurgical grade silicon** | 1 kg |
| **Solar grade silicon (UMG and polysilicon)** | 1 kg |
| **Multicrystalline silicon ingot** | 1 kg |
| **Multicrystalline silicon wafer** | 1 m² |
| **Al-BSF PV Cell** | 1 m² |
| **PV Module** | 1 m² |
| **BoS – Mounting structure (total site area)** | 1 m² |
| **BoS – Electrical installation (50MW)** | 1 p |
| **Installation (100MW)** | 1 kW$_{pk}$ |
| **Operation (during lifetime)** | 1 kWh$_e$ |

## S2 – Electricity Mixes

*Table 2. Share of different technologies in electricity production for China and Spain*

|  | China CN | Spain ES |
|---|---|---|
| **Coal** | 65,9% | 5,0% |
| **Oil** | 2,9% | 2,2% |
| **Natural gas** | 0,0% | 33,3% |
| **Nuclear** | 4,8% | 22,2% |
| **Hydropower** | 17,8% | 9,0% |
| **PV Solar** | 3,1% | 5,0% |
| **Wind** | 5,5% | 21,6% |
| **Cogeneration (biogas)** | 0,0% | 1,7% |

# S3 - Life Cycle Inventory tables

Values taken form ecoinvent 3.5 (Wernet et al., 2016), unless otherwise noted.

*Table 3 Metallurgical grade Silicon LCI*

| Item | Qty. | Unit | Source / comments |
|---|---|---|---|
| **Products** | | | |
| Silicon, metallurgical grade | 1 | kg | |
| **Materials** | | | |
| Charcoal {GLO}| market | Cut-off, U | 0,17 | kg | |
| Coke {GLO}| market | Cut-off, U | 23,12 | MJ | |
| Graphite {GLO}| market | Cut-off, U | 0,1 | kg | |
| Petroleum coke {GLO}| market | Cut-off, U | 0,5 | kg | |
| Oxygen, liquid {RER}| market | Cut-off, U | 0,02 | kg | |
| Silica sand {GLO}| market | Cut-off, U | 2,7 | kg | |
| Wood chips, wet, measured as dry mass {Europe w/o CH}| market | Cut-off, U | 0,549757 | kg | |
| Electric arc furnace converter {RER}| construction | Cut-off, U | 1E-11 | p | Updated for a metallurgical plant |
| **Electricity/heat** | | | |
| Electricity, high voltage, production | 11 | kWh | Selected electricity mix |
| **Emissions to air** | | | |
| Aluminum | 1,5508E-06 | kg | |
| Antimony | 7,8519E-09 | kg | |
| Arsenic | 9,4223E-09 | kg | |
| Boron | 2,7914E-07 | kg | |
| Cadmium | 3,1408E-10 | kg | |
| Calcium | 7,7538E-07 | kg | |
| Carbon dioxide, biogenic | 1,6098 | kg | |
| Carbon dioxide, fossil | 3,5808 | kg | |
| Carbon monoxide, biogenic | 0,00062027 | kg | |
| Carbon monoxide, fossil | 0,0013797 | kg | |
| Chlorine | 7,8519E-08 | kg | |
| Chromium | 7,8519E-09 | kg | |
| Cyanide | 6,8704E-06 | kg | |
| Fluorine | 3,8769E-08 | kg | |
| Hydrogen fluoride | 0,0005 | kg | |
| Hydrogen sulfide | 0,0005 | kg | |
| Iron | 3,8769E-06 | kg | |
| Lead | 3,4352E-07 | kg | |
| Mercury | 7,8519E-09 | kg | |
| Nitrogen oxides | 0,0097432 | kg | |
| NMVOC, non-methane volatile organic compounds, unspecified origin | 0,000096 | kg | |

| | | |
|---|---|---|
| Particulates, > 10 um | 0,0077538 | kg |
| Potassium | 0,00006203 | kg |
| Silicon | 0,0075134 | kg |
| Sodium | 7,7538E-07 | kg |
| Sulfur dioxide | 0,012241 | kg |
| Tin | 7,8519E-09 | kg |
| **Waste to treatment** | | |
| Slag from metallurgical grade silicon production {GLO}| market | Cut-off, U | 0,025 | kg |

*Table 4 Closed loop Siemens process for solar grade Silicon LCI*

| Item | Qty. | Unit | Source / comments |
|---|---|---|---|
| **Products** | | | |
| Silicon, solar grade, closed loop Siemens process | 1 | kg | From previous step |
| **Materials** | | | |
| Silicon, metallurgical grade (current study) | 1,2621 | kg | Estimation (Xie et al., 2018) (Fu et al., 2015) (Huang et al., 2017) |
| Chlorine, gaseous {RER}| market | Cut-off, U | 0,215 | kg | Updated for closed-loop process |
| Hydrogen, liquid {RER}| market | Cut-off, U | 0,0536 | kg | Updated for closed-loop process |
| Sodium hydroxide, w/o water, in 50% solution state {GLO}| market | Cut-off, U | 0,35 | kg | Final product etching |
| Quicklime, milled, packed {RER}| market quicklime, milled, packed | Cut-off, U | 0,58 | kg | Chlorine neutralization |
| Nitrogen, liquid {RER}| market | Cut-off, U | 12,9 | kg | Estimation for tank blanketing and systems purging |
| Water, completely softened, from decarbonised water, at user {RoW}| production | Cut-off, U | 1262 | kg | For cooling purposes |
| Graphite {GLO}| market | Cut-off, U | 0,0054 | kg | Estimation for seed chuks |
| Silicone factory {GLO}| market | Cut-off, U | 1E-11 | p | |
| **Electricity/heat** | | | |
| Electricity, high voltage, production market | 85 | kWh | Selected electricity mix Estimation (Huang et al., 2017; Woodhouse et al., 2019) |
| Heat, from steam, in chemical industry {RER}| market heat, from steam, in chemical industry | Cut-off, U | 180 | MJ | Estimation (Huang et al., 2017) |
| **Emissions to air** | | | |
| Heat, waste | 396 | MJ | |
| **Emissions to water** | | | |
| AOX, Adsorbable Organic Halogen as Cl | 1,2619E-05 | kg | |
| BOD5, Biological Oxygen Demand | 0,00020471 | kg | |
| COD, Chemical Oxygen Demand | 0,00202 | kg | |

| | | |
|---|---|---|
| Chloride | 0,035991 | kg |
| Copper | 1,0236E-07 | kg |
| Nitrogen | 0,00020751 | kg |
| Phosphate | 2,8043E-06 | kg |
| Sodium | 0,03379 | kg |
| Zinc | 1,963E-06 | kg |
| Iron | 5,6085E-06 | kg |
| DOC, Dissolved Organic Carbon | 0,00090998 | kg |
| TOC, Total Organic Carbon | 0,00090998 | kg |

*Table 5 Multicrystalline silicon ingot LCI*

| Item | Qty. | Unit | Source / comments |
|---|---|---|---|
| **Products** | | | |
| Silicon, multicrystalline ingot | 1 | kg | From previous step |
| **Materials** | | | |
| Silicon feedstock (current study) | 1.10 | kg | Selected silicon feedstock No mono scrap |
| Argon, liquid {RER}| market argon, liquid | Cut-off, U | 0.75 | kg | Higher demand for HPmc |
| Nitrogen, liquid {RER}| market | Cut-off, U | 0.0302 | kg | Helium is not used |
| Ceramic tile {GLO}| market | Cut-off, U | 0.164 | kg | |
| Graphite {GLO}| market | Cut-off, U | 0.0068 | kg | Crucible lining |
| Silica fume, densified {GLO}| market | Cut-off, U | 0.000448 | kg | Crucible coating |
| Alkylbenzene sulfonate, linear, petrochemical {GLO}| market | Cut-off, U | 0.08 | kg | Cleaning starting material |
| Nitric acid, w/o water, in 50% solution state {RER}| Cut-off, U | 0.0562 | kg | Etching |
| Hydrogen fluoride {RER}| market hydrogen fluoride | Cut-off, U | 0.00466 | kg | Etching |
| Water, deionised, from tap water, at user {Europe w/o CH}| market Cut-off, U | 3.2 | kg | |
| Nickel, 99.5% {GLO}| market | Cut-off, U | 0.000189 | kg | Wire for briquetting and cropping |
| Steel, chromium steel 18/8, hot rolled {GLO}| market | Cut-off, U | 0.001457 | kg | |
| Water, cooling, unspecified natural origin | 0.943 | m3 | |
| **Electricity/heat** | | | |
| Electricity, medium voltage | 11.35 | kWh | Higher demand for HPmc Selected mix |
| **Emissions to air** | | | |
| Heat, waste | 69.463 | MJ | |

*Table 6 Multicrystalline silicon wafer LCI*

| Item | Qty. | Unit | Source / comments |
|---|---|---|---|
| **Products** | | | |
| Silicon, multicrystalline wafer | 1 | m2 | From previous step |
| **Materials** | | | |
| Silicon, multi-Si, (current study) | 0.665 | kg | |
| Water, deionised, from tap, at user {RER w/o CH} |Cut-off, U | 57.81 | kg | Wafer cleaning |
| Tap water {Europe w/o CH}| market | Cut-off, U | 0.006 | kg | |
| Sodium hydroxide, w/o water, in 50% solution state {GLO}| market | Cut-off, U | 0.0038 | kg | |
| Acetic acid, w/o water, in 98% solution state {GLO}| market | Cut-off, U | 0.056 | kg | |
| Alkylbenzene sulfonate, linear, petrochemical {GLO}| market | Cut-off, U | 0.088 | kg | Detergent |
| Acrylic binder, w/o water, in 34% solution state {RER}| market acrylic binder, | Cut-off, U | 0.0028 | kg | |
| Flat glass, uncoated {GLO}| market | Cut-off, U | 0.0408 | kg | |
| Brass {RoW}| market brass | Cut-off, U | 0.00745 | kg | |
| Steel, chromium steel 18/8, hot rolled {GLO}| market | Cut-off, U | 0.0028 | kg | Diamond wire sawing |
| Steel, chromium steel 18/8, hot rolled {GLO}| market | Cut-off, U | 0.0177 | kg | |
| Wire drawing, steel {GLO}| market | Cut-off, U | 0.0028 | kg | |
| Nickel, 99.5% {GLO}| market | Cut-off, U | 0.0010 | kg | |
| Paper, woodfree, coated, at integrated mill/RER U | 0.19 | kg | Packaging |
| Polystyrene, high impact, HIPS, at plant/RER U | 0.014 | kg | |
| Packaging film, LDPE, at plant/RER U | 0.1 | kg | |
| Wafer plant/ES U | 6.7837E-09 | p | Updated |
| **Electricity/heat** | | | |
| Electricity, medium voltage | 12.13 | kWh | |
| Heat, district or industrial, natural gas {RER}| market | Cut-off, U | 4 | MJ | |
| **Emissions to air** | | | |
| Heat, waste | 74.9 | MJ | |
| **Emissions to water** | | | |
| AOX, Adsorbable Organic Halogen as Cl | 0.00050129 | kg | |
| Cadmium | 6.0508E-06 | kg | |
| Chromium | 3.0254E-05 | kg | |
| COD, Chemical Oxygen Demand | 0.029555 | kg | |
| Copper | 6.0508E-05 | kg | |
| Lead | 3.0254E-05 | kg | |
| Mercury | 6.0508E-06 | kg | |
| Nickel | 6.0508E-05 | kg | |
| Nitrogen | 0.0099449 | kg | |
| Phosphate | 0.00050129 | kg | |
| BOD5, Biological Oxygen Demand | 0.029555 | kg | |

| Item | Qty. | Unit | |
|---|---|---|---|
| DOC, Dissolved Organic Carbon | 0.011083 | kg | |
| TOC, Total Organic Carbon | 0.011083 | kg | |

*Table 7 Multicrystalline silicon solar cell LCI*

| Item | Qty. | Unit | Source / comments |
|---|---|---|---|
| **Products** | | | |
| Multicrystalline silicon PV cell (current study) | 1 | m2 | From previous step |
| **Materials** | | | |
| Silicon, multicrystalline wafer | 1.02 | m2 | |
| Nitrogen, liquid {RER}| market | Cut-off, U | 1.9273 | kg | Argon is no used |
| Oxygen, liquid {RER}| market | Cut-off, U | 0.0115 | kg | (Frischknecht et al., 2015; Huang et al., 2017) Industrial Tier 1 Producers |
| Ammonia, liquid {RER}| market | Cut-off, U | 0.0334 | kg | (Frischknecht et al., 2015; Huang et al., 2017) Industrial Tier 1 Producers |
| Calcium chloride {RER}| market calcium chloride | Cut-off, U | 0.021573 | kg | |
| Ethanol, w/o water, in 99.7% solution state, from ethylene {RER}| market | Cut-off, U | 0.019 | kg | Cleaning (Fu et al., 2015; Huang et al., 2017; Yu et al., 2017) |
| Water, completely softened, from decarbonised water, at user {RER}| production | Cut-off, U | 59 | kg | Industrial Tier 1 Producers |
| Nitric acid, w/o water, in 50% solution state {RER}| | Cut-off, U | 0.02416 | kg | (Huang et al., 2017; Luo et al., 2018; Yu et al., 2017) |
| Silicon tetrahydride {GLO}| market | Cut-off, U | 0.014771 | kg | (Huang et al., 2017; Yu et al., 2017) Industrial Tier 1 Producers |
| Hydrochloric acid, w/o water, in 30% solution state {RER}| market | Cut-off, U | 0.0785 | kg | (Frischknecht et al., 2015; Huang et al., 2017; Yu et al., 2017) |
| Hydrogen fluoride {RER}| market hydrogen fluoride | Cut-off, U | 0.01 | kg | (Frischknecht et al., 2015; Huang et al., 2017; Yu et al., 2017) |
| Metallization paste, back side, aluminium {RER}| market | Cut-off, U | 0.05495 | kg | Updated, Industrial Tier 1 Producers |
| Metallization paste, back side {RER}| market metallization paste, back side | Cut-off, U | 0.00081 | kg | Updated, Industrial Tier 1 Producers |
| Metallization paste, front side {RER}| market | Cut-off, U | 0.00366 | kg | Updated, Industrial Tier 1 Producers |
| Phosphoryl chloride {RER}| market | Cut-off, U | 0.0011 | kg | Emitter (Frischknecht et al., 2015; Huang et al., 2017; Luo et al., 2018; Yu et al., 2017) |
| Potassium hydroxide {GLO}| market | Cut-off, U | 0.0291 | kg | (Frischknecht et al., 2015; Huang et al., 2017; Luo et al., 2018; Yu et al., 2017) |
| Sodium silicate, spray powder, 80% {RER}| market sodium silicate, spray powder, 80% | Cut-off, U | 0.074786 | kg | |
| Solvent, organic {GLO}| market | Cut-off, U | 0.0014341 | kg | |

| | | | |
|---|---|---|---|
| Tetrafluoroethylene {GLO}| market | Cut-off, U | 0.0031558 | kg | |
| Steel, chromium steel 18/8, hot rolled {GLO}| market | Cut-off, U | 0.0000156 | kg | |
| Wire drawing, steel {GLO}| market | Cut-off, U | 0.0000156 | kg | |
| Polystyrene, expandable {GLO}| market | Cut-off, U | 0.00040722 | kg | |
| Cell plant | 0.0000004 | p | Updated current study |
| **Electricity/heat** | | | |
| Electricity, medium voltage | 25.75 | kWh | (Fu et al., 2015; Huang et al., 2017; Woodhouse et al., 2019) |
| Heat, district or industrial, natural gas {RER}| market group for | Cut-off, U | 3.5841 | MJ | (Fu et al., 2015) |
| **Emissions to air** | | | |
| Heat, waste | 108.88 | MJ | |
| Aluminium | 0.00077252 | kg | |
| Ethane, hexafluoro-, HFC-116 | 0.00011861 | kg | |
| Hydrogen chloride | 0.00071 | kg | |
| Hydrogen fluoride | 0.00057 | kg | |
| Lead | 0.00077252 | kg | |
| NMVOC, non-methane volatile organic compounds, unspecified origin | 0.19354 | kg | |
| Nitrogen oxides | 0.00686 | kg | |
| Methane, tetrafluoro-, CFC-14 | 0.00024763 | kg | |
| Particulates, < 2.5 um | 0.0026627 | kg | |
| Silicon | 7.2732E-05 | kg | |
| Silver | 0.00077252 | kg | |
| Sodium | 4.8488E-05 | kg | |
| Tin | 0.00077252 | kg | |
| Chlorine | 0.0031 | kg | |
| Ammonia | 0.0016 | kg | |
| VOC, volatile organic compounds | 0.0028492 | kg | |
| Emissions to water | | | |
| COD, Chemical Oxygen Demand | 0.004884 | kg | |
| Chloride | 0.0252 | kg | |
| Fluoride | 0.3187 | kg | |
| Waste to treatment | | | |
| Treatment, PV cell production effluent, to wastewater treatment, class 3/CH U | 0.0272 | m3 | |
| Disposal, waste, Si waferprod., inorg, 9.4% water, to residual material landfill/CH U | 0.27572 | kg | |

*Table 8 Multicrystalline silicon PV module LCI*

| Item | Qty. | Unit | Source / comments |
|---|---|---|---|
| **Products** | | | |
| Multicrystalline silicon PV module | 1 | m2 | |
| **Materials** | | | |
| Photovoltaic cell, multi-Si (current study) | 0.9059 | m2 | From previous step |
| Aluminium, wrought alloy {GLO}| market | Cut-off, U | 1.374 | kg | Frame Updated Industrial Tier 1 producers |
| Silicone product {RER}| market silicone product | Cut-off, U | 0.1977 | kg | Industrial Tier 1 producers |
| Tap water {Europe w/o CH}| market | Cut-off | 21.286 | kg | |
| Solar glass, low-iron {GLO}| market | Cut-off, U | 7.768 | kg | For 3.2 mm |
| Ethylvinylacetate, foil {GLO}| market | Cut-off, U | 0.88 | kg | Encapsulant updated Industrial Tier 1 producers |
| Polyvinylfluoride, film {GLO}| market | Cut-off, U | 0.4438 | kg | Backsheet updated Industrial Tier 1 producers |
| Wire drawing, copper {GLO}| market | Cut-off, U | 0.135 | kg | |
| Brazing solder, cadmium free {GLO}| market for | Cut-off, U | 0.0087647 | kg | Tab-ribbon |
| Tin {GLO}| market | Cut-off, U | 0.0236 | kg | |
| Copper {GLO}| market | Cut-off, U | 0.137 | kg | |
| Polyethylene terephthalate, granulate, amorphous {GLO}| market | Cut-off, U | 0.37297 | kg | |
| 1-propanol {GLO}| market | Cut-off, U | 0.004 | kg | |
| Acetone, liquid {RER}| market acetone, liquid | Cut-off, U | 0.012959 | kg | |
| Methanol {GLO}| market | Cut-off, U | 0.0021556 | kg | |
| Vinyl acetate {GLO}| market | Cut-off, U | 0.0016434 | kg | |
| Lubricating oil {RER}| market lubricating oil | Cut-off, U | 0.0016069 | kg | |
| Ethylene vinyl acetate copolymer {RER}| market ethylene vinyl acetate copolymer | Cut-off, U | 0.02228 | kg | Electric connectors |
| Copper {GLO}| market | Cut-off, U | 0.0396 | kg | Electric connectors |
| Glass fibre reinforced plastic, polyamide, injection moulded {GLO}| market | Cut-off, U | 0.18781 | kg | Connection box |
| Silicon capacitor (diode) | 0.002505 | kg | Updated |
| Corrugated board box {RER}| market | Cut-off, U | 1.0956 | kg | Packaging |
| Transport, lorry >16t, fleet average/RER U | 1.6093 | tkm | |
| Transport, freight, rail/RER U | 9.4484 | tkm | |
| Module plant | 6.5923E-09 | p | Updated Estimation |
| **Electricity/heat** | | | |
| Electricity, medium voltage, Aurinka 2019 market | 1.41 | kWh | |
| Heat, central or small-scale, natural gas {Europe w/o CH}| market | Cut-off, U | 5.4071 | MJ | |
| **Emissions to air** | | | |
| Heat, waste | 16.958 | MJ | |
| VOC, volatile organic compounds, unspecified origin | 0.00030609 | kg | (Yu et al., 2017) |
| **Waste to treatment** | | | |

| Disposal, municipal solid waste, 22.9% water, to municipal incineration/CH U | 0.0089 | kg |
| Disposal, polyvinylfluoride, 0.2% water, to municipal incineration/CH U | 0.1104 | kg |
| Disposal, plastics, mixture, 15.3% water, to municipal incineration/CH U | 1.6861 | kg |
| Disposal, used mineral oil, 10% water, to hazardous waste incineration/CH U | 0.0016069 | kg |
| Treatment, sewage, from residence, to wastewater treatment, class 2/CH U | 0.001616 | m3 |

*Table 9 Balance of system: Electric installation LCI*

| Item | Qty. | Unit | Source / comments |
| --- | --- | --- | --- |
| **Products** | | | |
| Electric installation for 50 MWp open ground module | 1 | p | |
| **Materials** | | | |
| Aluminium, wrought alloy {GLO}| market | Cut-off, U | 96945 | kg | Electrical cables |
| Wire drawing, copper {GLO}| market | Cut-off, U | 96945 | kg | |
| Polyethylene, high density, granulate {GLO}| Cut-off, U | 40535 | kg | |
| Polyvinylchloride, bulk polymerised {GLO}| market | Cut-off, U | 28083 | kg | |
| Epoxy resin, liquid {RER}| market | Cut-off, U | 2 | kg | |
| Epoxy resin, liquid {RoW}| market | Cut-off, U | 10 | kg | |
| Nylon 6 {GLO}| market | Cut-off, U | 1568.9 | kg | |
| Polycarbonate {GLO}| market | Cut-off, U | 13 | kg | Cabinets |
| **Waste to treatment** | | | |
| Waste electric wiring {RoW}| market | Cut-off, U | 165563 | kg | |
| Waste polyethylene/polypropylene product {Europe w/o CH}|| Cut-off, U | 40535 | kg | |
| Waste polyvinylchloride {RER}| market | Cut-off, U | 28083 | kg | |

Data from TINOSA project

*Table 10 Balance of system: Mounting system LCI*

| Item | Qty. | Unit | Source / comments |
| --- | --- | --- | --- |
| **Products** | | | |
| Photovoltaic mounting system, for 50MWp open ground | 1 | m2 | |
| **Resources** | | | |
| Transformation, from pasture, man made | 4.7 | m2 | 1.5 Ha / MWp |
| Transformation, to industrial area | 4.7 | m2 | |
| **Materials** | | | |
| Concrete, normal {RoW}| market | Cut-off, U | 0.00002 | m3 | |
| Reinforcing steel {GLO}| market | Cut-off, U | 1.4043 | kg | |

| Item | Qty. | Unit | Source / comments |
|---|---|---|---|
| Section bar rolling, steel {GLO}| market | Cut-off, U | 1.4087 | kg | |
| Steel, chromium steel 18/8, hot rolled {GLO}| market | Cut-off, U | 0.0044 | kg | Screws |
| Zinc coat, pieces {GLO}| market | Cut-off, U | 0.0312 | m2 | Galvanized steel |
| Polyethylene, high density, granulate {GLO}| market | Cut-off, U | 0.00090909 | kg | |
| Polystyrene, high impact {GLO}| market | Cut-off, U | 0.0045455 | kg | Cabinets |
| Waste paperboard, unsorted {GLO}| waste paperboard, unsorted, Recycled Content cut-off | Cut-off, U | -0.086364 | kg | Packaging |
| Corrugated board box {CA-QC}| market | Cut-off, U | 0.00079731 | kg | |
| Corrugated board box {RER}| market | Cut-off, U | 0.01783257 | kg | |
| Corrugated board box {RoW}| market | Cut-off, U | 0.06773413 | kg | |
| **Waste to treatment** | | | |
| Scrap steel {RoW}| market scrap steel | Cut-off, U | 1.4087 | kg | |
| Waste polyethylene/polypropylene product {Europe w/o CH}| market | Cut-off, U | 0.00090909 | kg | |
| Waste polystyrene isolation, flame-retardant {Europe w/o CH}| market | Cut-off, U | 0.0045455 | kg | |
| Waste reinforced concrete {Europe w/o CH}| market waste reinforced concrete | Cut-off, U | 0.00002 | kg | |

Data from TINOSA project

*Table 11 PV Installation LCI*

| Item | Qty. | Unit | Source / comments |
|---|---|---|---|
| **Products** | | | |
| PV installation | 100000 | kWpk | |
| **Resources** | | | |
| Occupation, unspecified, natural (non-use) | 47934884.9 | m2a | Calculated for polysilicon |
| | 48054550.6 | m2a | Calculated for UMG |
| **Materials** | | | |
| Photovoltaic panel, multi-Si, at plant/RER/I U AURINKA | 590829 | m2 | Calculated for polysilicon |
| | 594818 | m2 | Calculated for UMG |
| Inverter, 500kW {RER}| production | Cut-off, U | 273.4 | p | Calculated for mass ratio per kW of 1.75, assuming one full substitution. |
| Photovoltaic plant, electric installation for 50 MWp open ground | 2 | p | From previous step |
| Photovoltaic mounting system, for 50MWp open ground module {GLO}| production | Cut-off, U Aurinka | 1597829 | | From previous step Calculated for polysilicon |
| | 1601818 | m2 | From previous step Calculated for UMG-silicon |
| Diesel, burned in building machine {GLO}| market | Cut-off, U | 807684.211 | MJ | For construction (Antonanzas et al., 2019) |
| **Electricity/heat** | | | |
| Electricity, low voltage | 3792.94737 | kWh | For ES mix |

*Table 12 End-user: Electricity production LCI*

| Item | Qty. | Unit | Source / comments |
|---|---|---|---|
| **Products** | | | |
| Electricity PV production (100MW plant, 30 years) | 5168104629 | kWh | Calculated for production site |
| **Materials** | | | |
| PV Installation | 100000 | kWpk | From previous step |

# S4 - Life Cycle Assessment Results

*Table 13 Raw materials LCA results*

| Impact | Unit | Silica Sand (1 kg) | Metallurgical grade Silicon (1kg) | |
|---|---|---|---|---|
| | | | MG-Si ES | MG-Si CN |
| **CC** | kg CO2 eq | 2.13E-02 | 10.62 | 13.29 |
| **OD** | kg CFC11 eq | 3.77E-09 | 7.37E-07 | 2.83E-07 |
| **IR** | kBq U-235 eq | 2.61E-03 | 3.32E+00 | 8.05E-01 |
| **POF** | kg NMVOC eq | 5.69E-05 | 4.06E-02 | 6.30E-02 |
| **RI** | disease inc. | 6.80E-10 | 2.51E-07 | 3.03E-07 |
| **NCHHE** | CTUh | 2.93E-09 | 4.70E-07 | 8.08E-07 |
| **CHHE** | CTUh | 1.23E-10 | 3.75E-08 | 7.41E-08 |
| **ATF** | mol H+ eq | 7.17E-05 | 9.43E-02 | 2.22E-01 |
| **EF** | kg P eq | 1.15E-06 | 3.55E-03 | 1.10E-02 |
| **EM** | kg N eq | 1.60E-05 | 1.35E-02 | 1.83E-02 |
| **ET** | mol N eq | 1.79E-04 | 1.42E-01 | 1.78E-01 |
| **ETF** | CTUe | 9.17E-03 | 1.58 | 2.56 |
| **LU** | Pt | 8.14E-02 | 8.35 | 227.62 |
| **WS** | m3 depriv. | 1.38 | 1991 | 2842 |
| **RUEC** | MJ | 3.09E-01 | 161.10 | 116.51 |
| **RUMM** | kg Sb eq | 8.63E-09 | 1.06E-06 | 3.44E-06 |
| **CCF** | kg CO2 eq | 2.13E-02 | 10.36 | 13.05 |
| **CCB** | kg CO2 eq | 2.88E-06 | 2.51E-01 | 2.41E-01 |
| **CCLUT** | kg CO2 eq | 1.69E-07 | 7.54E-04 | 1.15E-03 |

CC: Climate change; OD: Ozone depletion; IR: Ionising radiation, HH; POF: Photochemical ozone formation, HH; RI: Respiratory inorganics; NCHHE: Non-cancer human health effects; CHHE: Cancer human health effects; ATF: Acidification terrestrial and freshwater; EF: Eutrophication freshwater; EM: Eutrophication marine; ET: Eutrophication terrestrial ; ETF: Ecotoxicity freshwater; LU: Land use; WS: Water scarcity; RUEC: Resource use, energy carriers; RUMM: Resource use, mineral and metals; CCF: Climate change – fossil; CCB: Climate change – biogenic; CCLUT: Climate change - land use and transform.

*Table 14 Silicon Feedstock: polysilicon and UMG-Silicon LCA results*

| Silicon Feedstock (1kg) | | | | | |
|---|---|---|---|---|---|
| Impact | Unit | Poly ES | Poly CN | UMG ES | UMG CN |
| **CC** | kg CO2 eq | 55.60 | 17.90 | 106.25 | 37.00 |
| **OD** | kg CFC11 eq | 6.58E-06 | 1.67E-06 | 4.19E-06 | 7.63E-07 |
| **IR** | kBq U-235 eq | 2.99E+01 | 1.06E+01 | 9.58E+00 | 2.94E+00 |
| **POF** | kg NMVOC eq | 1.33E-01 | 5.86E-02 | 4.26E-01 | 1.69E-01 |
| **RI** | disease inc. | 1.82E-06 | 6.22E-07 | 2.41E-06 | 8.44E-07 |
| **NCHHE** | CTUh | 2.54E-06 | 7.89E-07 | 7.95E-06 | 2.83E-06 |
| **CHHE** | CTUh | 2.44E-07 | 9.14E-08 | 6.97E-07 | 2.62E-07 |
| **ATF** | mol H+ eq | 2.99E-01 | 1.12E-01 | 1.85E+00 | 6.98E-01 |
| **EF** | kg P eq | 1.44E-02 | 5.12E-03 | 9.74E-02 | 3.64E-02 |
| **EM** | kg N eq | 4.12E-02 | 1.66E-02 | 1.38E-01 | 5.29E-02 |
| **ET** | mol N eq | 4.57E-01 | 1.74E-01 | 1.34E+00 | 5.07E-01 |
| **ETF** | CTUe | 9.73E+00 | 3.30E+00 | 2.44E+01 | 8.82E+00 |
| **LU** | Pt | 3.77E+02 | 2.98E+02 | 4.20E+02 | 3.14E+02 |
| **WS** | m3 depriv. | 1.31E+04 | 4.92E+03 | 2.56E+04 | 9.63E+03 |
| **RUEC** | MJ | 1.00E+03 | 3.01E+02 | 1.16E+03 | 3.60E+02 |
| **RUMM** | kg Sb eq | 4.88E-05 | 2.62E-05 | 3.39E-05 | 2.05E-05 |
| **CCF** | kg CO2 eq | 5.51E+01 | 1.75E+01 | 1.06E+02 | 3.67E+01 |
| **CCB** | kg CO2 eq | 4.57E-01 | 3.48E-01 | 3.87E-01 | 3.22E-01 |
| **CCLUT** | kg CO2 eq | 1.21E-02 | 4.02E-03 | 1.21E-02 | 4.03E-03 |

CC: Climate change; OD: Ozone depletion; IR: Ionising radiation, HH; POF: Photochemical ozone formation, HH; RI: Respiratory inorganics; NCHHE: Non-cancer human health effects; CHHE: Cancer human health effects; ATF: Acidification terrestrial and freshwater; EF: Eutrophication freshwater; EM: Eutrophication marine; ET: Eutrophication terrestrial ; ETF: Ecotoxicity freshwater; LU: Land use; WS: Water scarcity; RUEC: Resource use, energy carriers; RUMM: Resource use, mineral and metals; CCF: Climate change – fossil; CCB: Climate change – biogenic; CCLUT: Climate change - land use and transform.

*Table 15 Silicon Ingot LCA*

| mc-Si Ingot (1kg) | | | | | |
|---|---|---|---|---|---|
| Impact | Unit | Poly ES | Poly CN | UMG ES | UMG CN |
| **CC** | kg CO2 eq | 65.83 | 24.36 | 127.37 | 51.20 |
| **OD** | kg CFC11 eq | 7.76E-06 | 2.35E-06 | 4.85E-06 | 1.08E-06 |
| **IR** | kBq U-235 eq | 36.73 | 15.48 | 12.00 | 4.70 |
| **POF** | kg NMVOC eq | 1.58E-01 | 7.62E-02 | 5.14E-01 | 2.31E-01 |
| **RI** | disease inc. | 2.51E-06 | 1.20E-06 | 3.23E-06 | 1.51E-06 |
| **NCHHE** | CTUh | 3.18E-06 | 1.25E-06 | 9.75E-06 | 4.11E-06 |
| **CHHE** | CTUh | 3.21E-07 | 1.53E-07 | 8.71E-07 | 3.93E-07 |
| **ATF** | mol H+ eq | 3.61E-01 | 1.56E-01 | 2.25E+00 | 9.79E-01 |
| **EF** | kg P eq | 1.81E-02 | 7.89E-03 | 1.19E-01 | 5.18E-02 |
| **EM** | kg N eq | 4.95E-02 | 2.24E-02 | 1.67E-01 | 7.35E-02 |
| **ET** | mol N eq | 5.55E-01 | 2.44E-01 | 1.63E+00 | 7.11E-01 |
| **ETF** | CTUe | 12.24 | 5.16 | 30.03 | 12.92 |
| **LU** | Pt | 430.38 | 343.38 | 482.54 | 366.12 |
| **WS** | m3 depriv. | 15942.24 | 6950.64 | 31140.67 | 13578.73 |
| **RUEC** | MJ | 1207.60 | 435.94 | 1395.15 | 517.72 |
| **RUMM** | kg Sb eq | 7.03E-05 | 4.54E-05 | 5.22E-05 | 3.75E-05 |
| **CCF** | kg CO2 eq | 65.29 | 23.95 | 126.91 | 50.82 |
| **CCB** | kg CO2 eq | 5.27E-01 | 4.07E-01 | 4.42E-01 | 3.70E-01 |
| **CCLUT** | kg CO2 eq | 1.60E-02 | 7.18E-03 | 1.61E-02 | 7.20E-03 |

CC: Climate change; OD: Ozone depletion; IR: Ionising radiation, HH; POF: Photochemical ozone formation, HH; RI: Respiratory inorganics; NCHHE: Non-cancer human health effects; CHHE: Cancer human health effects; ATF: Acidification terrestrial and freshwater; EF: Eutrophication freshwater; EM: Eutrophication marine; ET: Eutrophication terrestrial ; ETF: Ecotoxicity freshwater; LU: Land use; WS: Water scarcity; RUEC: Resource use, energy carriers; RUMM: Resource use, mineral and metals; CCF: Climate change – fossil; CCB: Climate change – biogenic; CCLUT: Climate change - land use and transform.

*Table 16 Silicon wafer LCA*

| mc-Si Wafer (1m2) | | | | | |
|---|---|---|---|---|---|
| Impact | Unit | Poly ES | Poly CN | UMG ES | UMG CN |
| **CC** | kg CO2 eq | 48.59 | 21.01 | 95.73 | 45.08 |
| **OD** | kg CFC11 eq | 5.73E-06 | 2.13E-06 | 3.50E-06 | 9.91E-07 |
| **IR** | kBq U-235 eq | 28.00 | 13.87 | 9.05 | 4.20 |
| **POF** | kg NMVOC eq | 1.19E-01 | 6.46E-02 | 3.92E-01 | 2.04E-01 |
| **RI** | disease inc. | 1.91E-06 | 1.03E-06 | 2.46E-06 | 1.31E-06 |
| **NCHHE** | CTUh | 2.97E-06 | 1.69E-06 | 8.00E-06 | 4.25E-06 |
| **CHHE** | CTUh | 5.02E-07 | 3.91E-07 | 9.24E-07 | 6.06E-07 |
| **ATF** | mol H+ eq | 2.75E-01 | 1.38E-01 | 1.72E+00 | 8.77E-01 |
| **EF** | kg P eq | 1.44E-02 | 7.56E-03 | 9.16E-02 | 4.70E-02 |
| **EM** | kg N eq | 3.73E-02 | 1.93E-02 | 1.27E-01 | 6.51E-02 |
| **ET** | mol N eq | 4.16E-01 | 2.09E-01 | 1.24E+00 | 6.28E-01 |
| **ETF** | CTUe | 17.32 | 12.61 | 30.95 | 19.57 |
| **LU** | Pt | 333.40 | 275.54 | 373.36 | 295.94 |
| **WS** | m3 depriv. | 12344.28 | 6364.87 | 23988.74 | 12310.05 |
| **RUEC** | MJ | 912.55 | 399.39 | 1056.24 | 472.75 |
| **RUMM** | kg Sb eq | 1.01E-04 | 8.40E-05 | 8.67E-05 | 7.69E-05 |
| **CCF** | kg CO2 eq | 48.20 | 20.71 | 95.42 | 44.82 |
| **CCB** | kg CO2 eq | 3.72E-01 | 2.92E-01 | 3.07E-01 | 2.59E-01 |
| **CCLUT** | kg CO2 eq | 1.14E-02 | 5.56E-03 | 1.15E-02 | 5.57E-03 |

CC: Climate change; OD: Ozone depletion; IR: Ionising radiation, HH; POF: Photochemical ozone formation, HH; RI: Respiratory inorganics; NCHHE: Non-cancer human health effects; CHHE: Cancer human health effects; ATF: Acidification terrestrial and freshwater; EF: Eutrophication freshwater; EM: Eutrophication marine; ET: Eutrophication terrestrial ; ETF: Ecotoxicity freshwater; LU: Land use; WS: Water scarcity; RUEC: Resource use, energy carriers; RUMM: Resource use, mineral and metals; CCF: Climate change – fossil; CCB: Climate change – biogenic; CCLUT: Climate change - land use and transform.

*Table 17 PV Cell LCA results*

| mc-Si PV Cell (1 m²) | | | | | |
|---|---|---|---|---|---|
| Impact | Unit | Poly ES | Poly CN | UMG ES | UMG CN |
| **CC** | kg CO2 eq | 68,98 | 40,85 | 130,29 | 78,62 |
| **OD** | kg CFC11 eq | 1,84E-05 | 1,47E-05 | 1,55E-05 | 1,29E-05 |
| **IR** | kBq U-235 eq | 36,92 | 22,51 | 12,28 | 7,33 |
| **POF** | kg NMVOC eq | 3,71E-01 | 3,15E-01 | 7,25E-01 | 5,33E-01 |
| **RI** | disease inc. | 3,27E-06 | 2,38E-06 | 3,99E-06 | 2,82E-06 |
| **NCHHE** | CTUh | 3,14E-05 | 3,01E-05 | 3,80E-05 | 3,42E-05 |
| **CHHE** | CTUh | 6,96E-07 | 5,82E-07 | 1,24E-06 | 9,20E-07 |
| **ATF** | mol H+ eq | 3,91E-01 | 2,51E-01 | 2,27E+00 | 1,41E+00 |
| **EF** | kg P eq | 2,68E-02 | 1,99E-02 | 1,27E-01 | 8,17E-02 |
| **EM** | kg N eq | 6,15E-02 | 4,31E-02 | 1,78E-01 | 1,15E-01 |
| **ET** | mol N eq | 6,72E-01 | 4,61E-01 | 1,74E+00 | 1,12E+00 |
| **ETF** | CTUe | 87,17 | 82,37 | 104,89 | 93,29 |
| **LU** | Pt | 417,52 | 358,50 | 469,48 | 390,52 |
| **WS** | m3 depriv. | 16089,11 | 9990,11 | 31230,33 | 19318,06 |
| **RUEC** | MJ | 1212,30 | 688,88 | 1399,14 | 803,98 |
| **RUMM** | kg Sb eq | 1,96E-03 | 1,94E-03 | 1,94E-03 | 1,93E-03 |
| **CCF** | kg CO2 eq | 68,49 | 40,45 | 129,88 | 78,27 |
| **CCB** | kg CO2 eq | 4,66E-01 | 3,84E-01 | 3,81E-01 | 3,32E-01 |
| **CCLUT** | kg CO2 eq | 2,57E-02 | 1,97E-02 | 2,57E-02 | 1,97E-02 |

CC: Climate change; OD: Ozone depletion; IR: Ionising radiation, HH; POF: Photochemical ozone formation, HH; RI: Respiratory inorganics; NCHHE: Non-cancer human health effects; CHHE: Cancer human health effects; ATF: Acidification terrestrial and freshwater; EF: Eutrophication freshwater; EM: Eutrophication marine; ET: Eutrophication terrestrial ; ETF: Ecotoxicity freshwater; LU: Land use; WS: Water scarcity; RUEC: Resource use, energy carriers; RUMM: Resource use, mineral and metals; CCF: Climate change – fossil; CCB: Climate change – biogenic; CCLUT: Climate change - land use and transform.

*Table 18 PV Module LCA results*

| mc-Si PV Module (1m²) | | | | | |
|---|---|---|---|---|---|
| Impact | Unit | Poly ES | Poly CN | UMG ES | UMG CN |
| **CC** | kg CO2 eq | 108,44 | 82,95 | 164,70 | 117,89 |
| **OD** | kg CFC11 eq | 1,99E-05 | 1,65E-05 | 1,72E-05 | 1,49E-05 |
| **IR** | kBq U-235 eq | 36,66 | 23,60 | 14,05 | 9,56 |
| **POF** | kg NMVOC eq | 5,07E-01 | 4,57E-01 | 8,32E-01 | 6,58E-01 |
| **RI** | disease inc. | 6,06E-06 | 5,25E-06 | 6,72E-06 | 5,66E-06 |
| **NCHHE** | CTUh | 3,89E-05 | 3,77E-05 | 4,49E-05 | 4,14E-05 |
| **CHHE** | CTUh | 2,03E-06 | 1,93E-06 | 2,53E-06 | 2,24E-06 |
| **ATF** | mol H+ eq | 7,41E-01 | 6,14E-01 | 2,47E+00 | 1,69E+00 |
| **EF** | kg P eq | 5,40E-02 | 4,77E-02 | 1,46E-01 | 1,05E-01 |
| **EM** | kg N eq | 1,15E-01 | 9,84E-02 | 2,22E-01 | 1,65E-01 |
| **ET** | mol N eq | 1,23 | 1,04 | 2,21 | 1,65 |
| **ETF** | CTUe | 120,98 | 116,63 | 137,24 | 126,73 |
| **LU** | Pt | 692,92 | 639,46 | 740,61 | 669,07 |
| **WS** | m3 depriv. | 14879,40 | 9354,31 | 28774,54 | 17983,22 |
| **RUEC** | MJ | 1665,43 | 1191,26 | 1836,88 | 1297,73 |
| **RUMM** | kg Sb eq | 2,74E-03 | 2,73E-03 | 2,73E-03 | 2,72E-03 |
| **CCF** | kg CO2 eq | 107,71 | 82,31 | 164,05 | 117,29 |
| **CCB** | kg CO2 eq | 6,15E-01 | 5,41E-01 | 5,37E-01 | 4,93E-01 |
| **CCLUT** | kg CO2 eq | 1,11E-01 | 1,06E-01 | 1,11E-01 | 1,06E-01 |

CC: Climate change; OD: Ozone depletion; IR: Ionising radiation, HH; POF: Photochemical ozone formation, HH; RI: Respiratory inorganics; NCHHE: Non-cancer human health effects; CHHE: Cancer human health effects; ATF: Acidification terrestrial and freshwater; EF: Eutrophication freshwater; EM: Eutrophication marine; ET: Eutrophication terrestrial ; ETF: Ecotoxicity freshwater; LU: Land use; WS: Water scarcity; RUEC: Resource use, energy carriers; RUMM: Resource use, mineral and metals; CCF: Climate change – fossil; CCB: Climate change – biogenic; CCLUT: Climate change - land use and transform.

*Table 19 Balance of system LCA results*

| Balance of System | | | |
|---|---|---|---|
| Impact | Unit | Electrical Installation (1 kWpk) | Mounting Structure (1 m²) (*) |
| CC | kg CO2 eq | 35.76 | 3.70 |
| OD | kg CFC11 eq | 1.04E-06 | 2.32E-07 |
| IR | kBq U-235 eq | 9.11E-01 | 0.17 |
| POF | kg NMVOC eq | 1.04E-01 | 1.75E-02 |
| RI | disease inc. | 2.53E-06 | 2.94E-07 |
| NCHHE | CTUh | 6.79E-06 | 2.45E-06 |
| CHHE | CTUh | 1.37E-06 | 5.47E-07 |
| ATF | mol H+ eq | 2.11E-01 | 1.95E-02 |
| EF | kg P eq | 1.57E-02 | 2.19E-03 |
| EM | kg N eq | 3.27E-02 | 3.91E-03 |
| ET | mol N eq | 3.32E-01 | 0.04 |
| ETF | CTUe | 40.27 | 8.61 |
| LU | Pt | 93.86 | 621.78 |
| WS | m3 depriv. | 15.71 | 1.74 |
| RUEC | MJ | 340.88 | 39.72 |
| RUMM | kg Sb eq | 2.25E-04 | 2.39E-04 |
| CCF | kg CO2 eq | 35.56 | 3.69 |
| CCB | kg CO2 eq | 1.00E-01 | 5.25E-03 |
| CCLUT | kg CO2 eq | 9.62E-02 | 7.97E-03 |

CC: Climate change; OD: Ozone depletion; IR: Ionising radiation, HH; POF: Photochemical ozone formation, HH; RI: Respiratory inorganics; NCHHE: Non-cancer human health effects; CHHE: Cancer human health effects; ATF: Acidification terrestrial and freshwater; EF: Eutrophication freshwater; EM: Eutrophication marine; ET: Eutrophication terrestrial ; ETF: Ecotoxicity freshwater; LU: Land use; WS: Water scarcity; RUEC: Resource use, energy carriers; RUMM: Resource use, mineral and metals; CCF: Climate change – fossil; CCB: Climate change – biogenic; CCLUT: Climate change – land use and transform.

(*) For a 50MW open ground installation

*Table 20 PV Installation LCA results*

| PV Installation (1 kWp) | | | | | |
|---|---|---|---|---|---|
| Impact | Unit | Poly ES | Poly CN | UMG ES | UMG CN |
| **CC** | kg CO2 eq | 7,72E+02 | 6,25E+02 | 1,10E+03 | 8,33E+02 |
| **OD** | kg CFC11 eq | 1,26E-04 | 1,07E-04 | 1,10E-04 | 9,71E-05 |
| **IR** | kBq U-235 eq | 2,26E+02 | 1,50E+02 | 9,23E+01 | 6,62E+01 |
| **POF** | kg NMVOC eq | 3,61E+00 | 3,33E+00 | 5,53E+00 | 4,53E+00 |
| **RI** | disease inc. | 4,56E-05 | 4,11E-05 | 4,95E-05 | 4,35E-05 |
| **NCHHE** | CTUh | 3,12E-04 | 3,07E-04 | 3,48E-04 | 3,29E-04 |
| **CHHE** | CTUh | 2,55E-05 | 2,50E-05 | 2,84E-05 | 2,68E-05 |
| **ATF** | mol H+ eq | 5,50E+00 | 4,78E+00 | 1,57E+01 | 1,12E+01 |
| **EF** | kg P eq | 4,74E-01 | 4,38E-01 | 1,02E+00 | 7,79E-01 |
| **EM** | kg N eq | 8,39E-01 | 7,45E-01 | 1,47E+00 | 1,14E+00 |
| **ET** | mol N eq | 8,94E+00 | 7,86E+00 | 1,47E+01 | 1,15E+01 |
| **ETF** | CTUe | 1,10E+03 | 1,08E+03 | 1,20E+03 | 1,14E+03 |
| **LU** | Pt | 6,39E+04 | 6,37E+04 | 6,41E+04 | 6,39E+04 |
| **WS** | m3 depriv. | 8,80E+04 | 5,57E+04 | 1,70E+05 | 1,07E+05 |
| **RUEC** | MJ | 1,14E+04 | 8,60E+03 | 1,24E+04 | 9,23E+03 |
| **RUMM** | kg Sb eq | 2,25E-02 | 2,25E-02 | 2,24E-02 | 2,24E-02 |
| **CCF** | kg CO2 eq | 7,67E+02 | 6,21E+02 | 1,10E+03 | 8,29E+02 |
| **CCB** | kg CO2 eq | 3,97E+00 | 3,56E+00 | 3,51E+00 | 3,27E+00 |
| **CCLUT** | kg CO2 eq | 1,02E+00 | 9,96E-01 | 1,02E+00 | 9,96E-01 |

CC: Climate change; OD: Ozone depletion; IR: Ionising radiation, HH; POF: Photochemical ozone formation, HH; RI: Respiratory inorganics; NCHHE: Non-cancer human health effects; CHHE: Cancer human health effects; ATF: Acidification terrestrial and freshwater; EF: Eutrophication freshwater; EM: Eutrophication marine; ET: Eutrophication terrestrial ; ETF: Ecotoxicity freshwater; LU: Land use; WS: Water scarcity; RUEC: Resource use, energy carriers; RUMM: Resource use, mineral and metals; CCF: Climate change – fossil; CCB: Climate change – biogenic; CCLUT: Climate change - land use and transform.

*Table 21 PV Electricity production LCIA results*

| Electricity production (1 kWhe)(*) | | | | | |
|---|---|---|---|---|---|
| Impact | Unit | Poly ES | Poly CN | UMG ES | UMG CN |
| **CC** | kg CO2 eq | 1,49E-02 | 1,21E-02 | 2,14E-02 | 1,61E-02 |
| **OD** | kg CFC11 eq | 2,44E-09 | 2,07E-09 | 2,13E-09 | 1,88E-09 |
| **IR** | kBq U-235 eq | 4,37E-03 | 2,90E-03 | 1,79E-03 | 1,28E-03 |
| **POF** | kg NMVOC eq | 6,99E-05 | 6,45E-05 | 1,07E-04 | 8,77E-05 |
| **RI** | disease inc. | 8,83E-10 | 7,95E-10 | 9,58E-10 | 8,42E-10 |
| **NCHHE** | CTUh | 6,04E-09 | 5,94E-09 | 6,73E-09 | 6,36E-09 |
| **CHHE** | CTUh | 4,93E-10 | 4,83E-10 | 5,50E-10 | 5,19E-10 |
| **ATF** | mol H+ eq | 1,06E-04 | 9,25E-05 | 3,04E-04 | 2,16E-04 |
| **EF** | kg P eq | 9,16E-06 | 8,48E-06 | 1,97E-05 | 1,51E-05 |
| **EM** | kg N eq | 1,62E-05 | 1,44E-05 | 2,85E-05 | 2,21E-05 |
| **ET** | mol N eq | 1,73E-04 | 1,52E-04 | 2,85E-04 | 2,22E-04 |
| **ETF** | CTUe | 2,14E-02 | 2,10E-02 | 2,32E-02 | 2,21E-02 |
| **LU** | Pt | 1,24E+00 | 1,23E+00 | 1,24E+00 | 1,24E+00 |
| **WS** | m3 depriv. | 1,70E+00 | 1,08E+00 | 3,29E+00 | 2,07E+00 |
| **RUEC** | MJ | 2,20E-01 | 1,66E-01 | 2,39E-01 | 1,79E-01 |
| **RUMM** | kg Sb eq | 4,35E-07 | 4,35E-07 | 4,33E-07 | 4,34E-07 |
| **CCF** | kg CO2 eq | 1,49E-02 | 1,20E-02 | 2,13E-02 | 1,60E-02 |
| **CCB** | kg CO2 eq | 7,68E-05 | 6,88E-05 | 6,79E-05 | 6,33E-05 |
| **CCLUT** | kg CO2 eq | 1,98E-05 | 1,93E-05 | 1,98E-05 | 1,93E-05 |

CC: Climate change; OD: Ozone depletion; IR: Ionising radiation, HH; POF: Photochemical ozone formation, HH; RI: Respiratory inorganics; NCHHE: Non-cancer human health effects; CHHE: Cancer human health effects; ATF: Acidification terrestrial and freshwater; EF: Eutrophication freshwater; EM: Eutrophication marine; ET: Eutrophication terrestrial ; ETF: Ecotoxicity freshwater; LU: Land use; WS: Water scarcity; RUEC: Resource use, energy carriers; RUMM: Resource use, mineral and metals; CCF: Climate change – fossil; CCB: Climate change – biogenic; CCLUT: Climate change - land use and transform.

(*) The calculated total production for the 100MW site during its expected lifetime (30 years) is 5168.1 GWh

## S5 – LCA Results: Electricity Mix

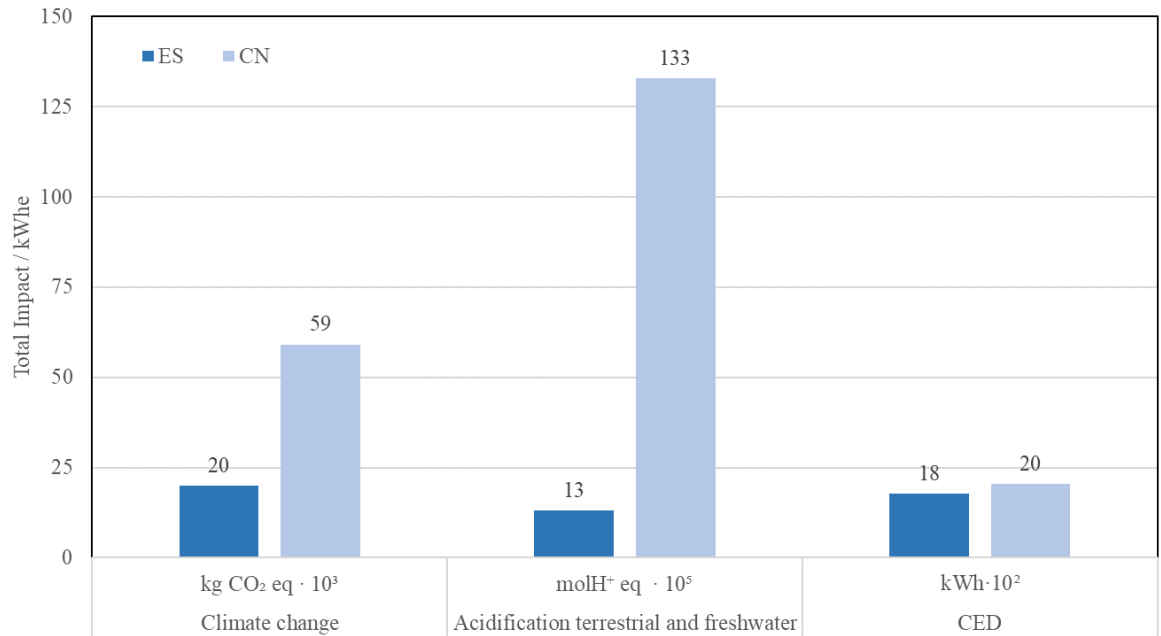

*Figure 1. Main impacts per kWh of the electricity mixes*

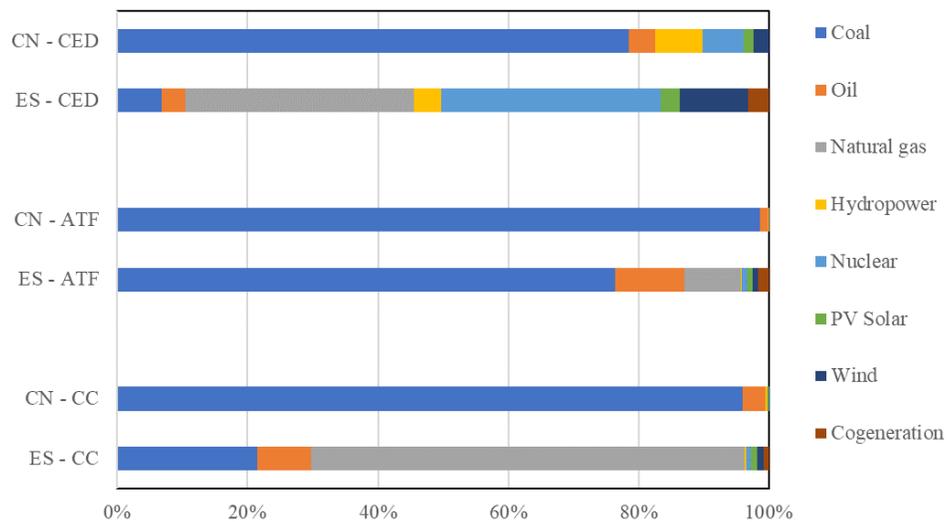

*Figure 2. Normalized technology contributions to CC, ATF and CED for ES an CN electricity mixes*

## S6 – LCA Results: Literature Summary

*Table 22. Summary of selected results from this work and literature for UMG and polysilicon PV electricity generation (per kWhe)*

| Silicon Feedstock | PV Cell / Module | PV System | Irradiation kWh/m²·year | Assessment method | Elec. mix | CC gCO$_{2eq}$ | CED kWh·10² | EPBT Years | Source |
|---|---|---|---|---|---|---|---|---|---|
| UMG mc-Si | BSF 325.9W 18.43% | 30 years 100MW Tilted fixed PR 82.5% | 2160 | EF | ES | 12.10 | 5.67 | 0.52 | Present work |
|  |  |  |  |  | CN | 16.12 | 5.94 | 0.55 |  |
| UMG | BSF 14.3% | 30 years Rooftop PR 75% | 1700 | IPCC2007 | NO | 27 | n.a. | 1.05 | (de Wild-Scholten and Gløckner, 2012) |
| UMG mc-Si | BSF 230W 14.08% | 30 years 10MW Tilted fixed PR 79.6% | 1670 | IPCC2007 | CN | 20 | 9.87 | 3.06 | (Yu et al., 2017) |
| Polysilicon mc-Si | BSF 328.1 W 18.55% | 30 years 100MW Tilted fixed PR 82.5% | 2160 | EF | ES | 14.95 | 7.50 | 0.69 | Present work |
|  |  |  |  |  | CN | 21.38 | 7.93 | 0.72 |  |
| Polysilicon mc-Si mono-Si | BSF 16,7% 18% | 30 years 100MW Tilted fixed PR 80% | 2300 | Midpoint CML | CN | 27 | nr 9.44 | 0.95 | (Raugei et al., 2020) |
| Polysilicon mc-Si | BSF 17.5% | 30 years Tilted Fixed PR 85% | 2327 | Midpoint ReCiPe 1.11 | CN | 23 | n.a. | 0.65 | (Antonanzas et al., 2019) |
| Polysilicon mc-Si | BSF 14.3% | 30 years Rooftop PR 75% | 1700 | IPCC2007 | UCTE | 31.3 | n.a. | 1.25 | (de Wild-Scholten and Gløckner, 2012) |
| Polysilicon mc-Si | BSF 15.9% PERC 16.7% | 25 years PR 78.5% | 1580 | IPCC 2013 | SIN | 30.2 29.2 | 8.61 8.78 | 1.11 1.08 | (Luo et al., 2018) |
| Polysilicon mc-Si | BSF 18.7% | 25 years PR 85.0% BOS excluded | 1770 | Midpoint ReCiPe 1.11 | CN | 21 | n.a. | 1.16 | (Jia et al., 2020) |
| Polysilicon mc-Si | BSF 16% | 25 years PR 85.0% BOS excluded | 2100 | CML2001 | CN | 51 | 14.1 | 2.22 | (Fu et al., 2015) |

mc: multicrystalline, mono; monocrystalline; electricity mixes: ES: Spain, CN: China, NO: Norway, SIN: Singapore, UCTE: Union for the Coordination of the Transmission of Electricity (nowadays ENTSO-E, Europe), SIN: Singapore; Cell technologies: BSF: Aluminium back surface field, PERC: Passivated Emitter and Rear Cell; PR: performance ratio; CC: climate change, CED: cumulative energy demand, EPBT: energy payback time; nr: non-renewable, r: renewable; n.a.: not available.

## S7 – LCA Results: PV Electricity contributions breakdown

*Table 23. Contribution of the different process step to Climate Change, CC (gCO$_{2eq}$ / kWh$_e$), Acidification terrestrial and freshwater, ATF ($10^{-5}$ molH$^+_{eq}$/ kWh$_e$), and Cumulative energy demand ($10^{-2}$ kWh/ kWh$_e$*

|  |  | Impact category | PROCESS STEP | | | | | |
|---|---|---|---|---|---|---|---|---|
|  |  |  | Silicon | Ingot | Wafer | PV Cell | Module | BoS&Inst. |
| UMG | ES | CC | 1,39 | 0,33 | 0,51 | 2,03 | 5,29 | 2,55 |
|  |  | ATF | 0,87 | 0,23 | 0,37 | 1,15 | 4,45 | 2,18 |
|  |  | CED | 0,88 | 0,26 | 0,43 | 1,05 | 2,12 | 0,94 |
|  | CN | CC | 2,88 | 0,74 | 1,17 | 3,40 | 5,37 | 2,55 |
|  |  | ATF | 5,43 | 1,49 | 2,40 | 5,38 | 4,71 | 2,18 |
|  |  | CED | 0,98 | 0,29 | 0,47 | 1,14 | 2,12 | 0,94 |
| poly | ES | CC | 4,30 | 0,33 | 0,51 | 2,01 | 5,25 | 2,55 |
|  |  | ATF | 2,31 | 0,23 | 0,37 | 1,14 | 4,42 | 2,18 |
|  |  | CED | 2,74 | 0,26 | 0,43 | 1,04 | 2,10 | 0,94 |
|  | CN | CC | 8,21 | 0,74 | 1,17 | 3,38 | 5,34 | 2,55 |
|  |  | ATF | 14,32 | 1,48 | 2,39 | 5,34 | 4,67 | 2,18 |
|  |  | CED | 2,99 | 0,29 | 0,47 | 1,13 | 2,11 | 0,94 |

## S8 – LCA Results: UMG-Si production stages contributions breakdown

*Table 24. Contribution of process steps in Climate Change for UMG (kg CO2eq/ kg of SoG-Si)*

|  | MG-Si | Slagging | Vacuum Refining | Directional Solidification | Other |
|---|---|---|---|---|---|
| **UMG ES** | 10.00 | 4.27 | 1.97 | 1.48 | 0.17 |
| **UMG CN** | 17.51 | 9.08 | 6.07 | 4.02 | 0.31 |

## S9 – Cumulative Energy Demand: PV Electricity total and background processes

Table 25. Cumulative Energy Demand detailed results for PV electricity generation and background processes for UMG and poly scenarios (kWh·106 / kWhe).

| CED kWh·10$^6$ / kWh$_e$ | Total | | | | Background processes | |
|---|---|---|---|---|---|---|
| | UMG ES | UMG CN | poly ES | poly CN | UMG | poly |
| **Non-renewable, fossil** | 39765 | 48379 | 51035 | 64813 | 3795 | 3944 |
| **Non-renewable, nuclear** | 9429 | 4756 | 13947 | 6473 | 404 | 467 |
| **Non-renewable, biomass** | 4 | 4 | 4 | 4 | 0 | 0 |
| **Renewable, biomass** | 2249 | 1746 | 2716 | 1910 | 185 | 178 |
| **Renewable, wind, solar, etc.** | 2500 | 967 | 3922 | 1469 | 14 | 19 |
| **Renewable, water** | 2798 | 3535 | 3406 | 4584 | 281 | 305 |
| **Subtotal, NRE** | 49198 | 53140 | 64986 | 71290 | 4199 | 4411 |
| **Subtotal, RE** | 7548 | 6248 | 10044 | 7964 | 480 | 502 |
| **Total** | 56746 | 59387 | 75030 | 79254 | 4679 | 4913 |